\documentclass[universe,article,accept,moreauthors,pdftex,trackchanges]{Definitions/mdpi}

\usepackage{subfigure}
\makeatletter
\renewcommand{\@thesubfigure}{\normalsize(\textbf{\alph{subfigure}})}
\makeatother

\firstpage{1} 
\pubvolume{8}
\issuenum{10}
\articlenumber{152}
\pubyear{2022}
\copyrightyear{2022}
\externaleditor{Academic Editors: Gang Zhao, Zi-Gao Dai and Da-Ming Wei}

\datereceived{8 August 2022}
\dateaccepted{27 September 2022}
\datepublished{1 October 2022}
\hreflink{https://doi.org/10.3390/\linebreak universe8100512} 

\Title{A Comprehensive Study of Bright Fermi-GBM Short Gamma-Ray Bursts: II. Very Short Burst and Its Implications}

\TitleCitation{A Comprehensive Study of Bright Fermi-GBM Short Gamma-Ray Bursts: II. Very Short Burst and Its Implications}

\Author{Ying-Yong Hu, Yao-Lin Huang, Jia-Wei Huang, Zan Zhu and Qing-Wen Tang *\orcidA{}}

\AuthorNames{Ying-Yong Hu, Yao-Lin Huang, Jia-Wei Huang, Zan Zhu and Qing-Wen Tang}

\AuthorCitation{Hu, Y.-Y.; Huang, Y.-L.; Huang, J.-W; Zhu, Z.; Tang, Q.-W.}

\address[1]{	Department of Physics, Nanchang University, Nanchang 330031, China}

\corres{\hangafter=1 \hangindent=1.05em \hspace{-0.82em}Correspondence: qwtang@ncu.edu.cn}

\abstract{A thermal component is suggested to be the physical composition of the ejecta of several bright gamma-ray bursts (GRBs). Such a thermal component is discovered in the time-integrated spectra of several short GRBs as well as long GRBs. In this work, we present a comprehensive analysis of ten very short GRBs detected by Fermi Gamma-Ray Burst Monitor to search for the thermal component. We found that both the resultant low-energy spectral index and the peak energy in each GRB imply a common hard spectral feature, which is in favor of the main classification of the short/hard versus long/soft dichotomy in the GRB duration. We also found moderate evidence for the detection of thermal component in eight GRBs. Although such a thermal component contributes a small proportion of the global prompt gamma-ray emission, the modified thermal-radiation mechanism could enhance the proportion significantly, such as in subphotospheric dissipation.}


\begin{document}
	
	\section{Introduction}
	
	Gamma-ray Burst (GRB) is the most intense transient astrophysical phenomena in the universe. Long GRB (LGRB, duration longer than 2 s) is believed to origin from the core-collapse of a massive star~\citep{Woosley1993, Paczynski1998,Galama1998,Stanek2003,Woosley2006,Fruchter2006}. However, short GRB (SGRB, duration less than 2 s) shares a distinct origin, such as a merger process of two compact objects, e.g., two neutron stars (NS-NS), which is first proved by a gravitational-wave GRB, GRB 170817A~\citep{Abbott2017}. For both types of GRBs, the dominate component of the gamma-ray emission at the photospheric radius, where energy dissipation takes place in the optically thick regime, is the so-called quisa-thermal component, which gradually fades across the burst duration. This thermal component represented by a standard blackbody function is discovered in several bright GRBs, such as GRB 090902B, GRB 120323A and GRB 170206A~\citep{Abdo2009,Guiriec2013,Tang2021,Zhao2022a,Zhao2022b}. Some modified blackbody modesl to reproduce the thermal photospheric emission are proposed and employed to fit the spectra of several GRBs~\citep{Arimoto2016,Lv2017,Hou2018,Iyyani2021}. Subphotospheric emission is also a popular mechanism for the thermal component in the Fermi era~\citep{Ahlgren2015,Beloborodov2017,Ryde2019}. Photospheric emission from a structured jet or a hybrid relativistic outflow is invoked in some GRBs~\citep{Lundman2013,Meng2019,Meng2022}. Photospheric emission models via the Compton scattering are also proposed to broaden the thermal peak~\citep{Lazzati2010,Giannios2012,Vyas2021a,Vyas2021b}.
	
	In the fireball model, GRB photosphere often occurs in the early phase, after which there is the emission region dominated by the internal shock, thus the thermal component and the non-thermal component would dominate the different emission phase~\citep{Guiriec2015}. The spectral evolution that includes the thermal and non-thermal component is also confirmed in several GRBs~\citep{Guiriec2010,Axelsson2012,Burgess2014,Yu2019}. In this work, we select a GRB sample with a very short duration detected by  Fermi Gamma-Ray Burst Monitor (Fermi/GBM) to judge these features, such with a burst duration shorter than 0.05 s, which often cannot be performed the time-resolved spectral analysis. In these GRBs, the thermal component would dominate prompt gamma rays if the flux of the thermal emission exceeds that of the non-thermal emission. In Section~\ref{method}, we present the sample selection and describe the method of spectral fitting. The results for four spectral models are presented in Section~\ref{results}. A short discussion is given in~ Section~\ref{discussion}. We present the summary and conclusions in Section~\ref{conclusion}.
	
	\section{Data Reduction \label{method}}
	
	\subsection{Sample Selection}
\begin{figure}[H]

\begin{adjustwidth}{-\extralength}{0cm}
		\centering
		\subfigure[~090802]{
			\includegraphics[width=0.55\textwidth]{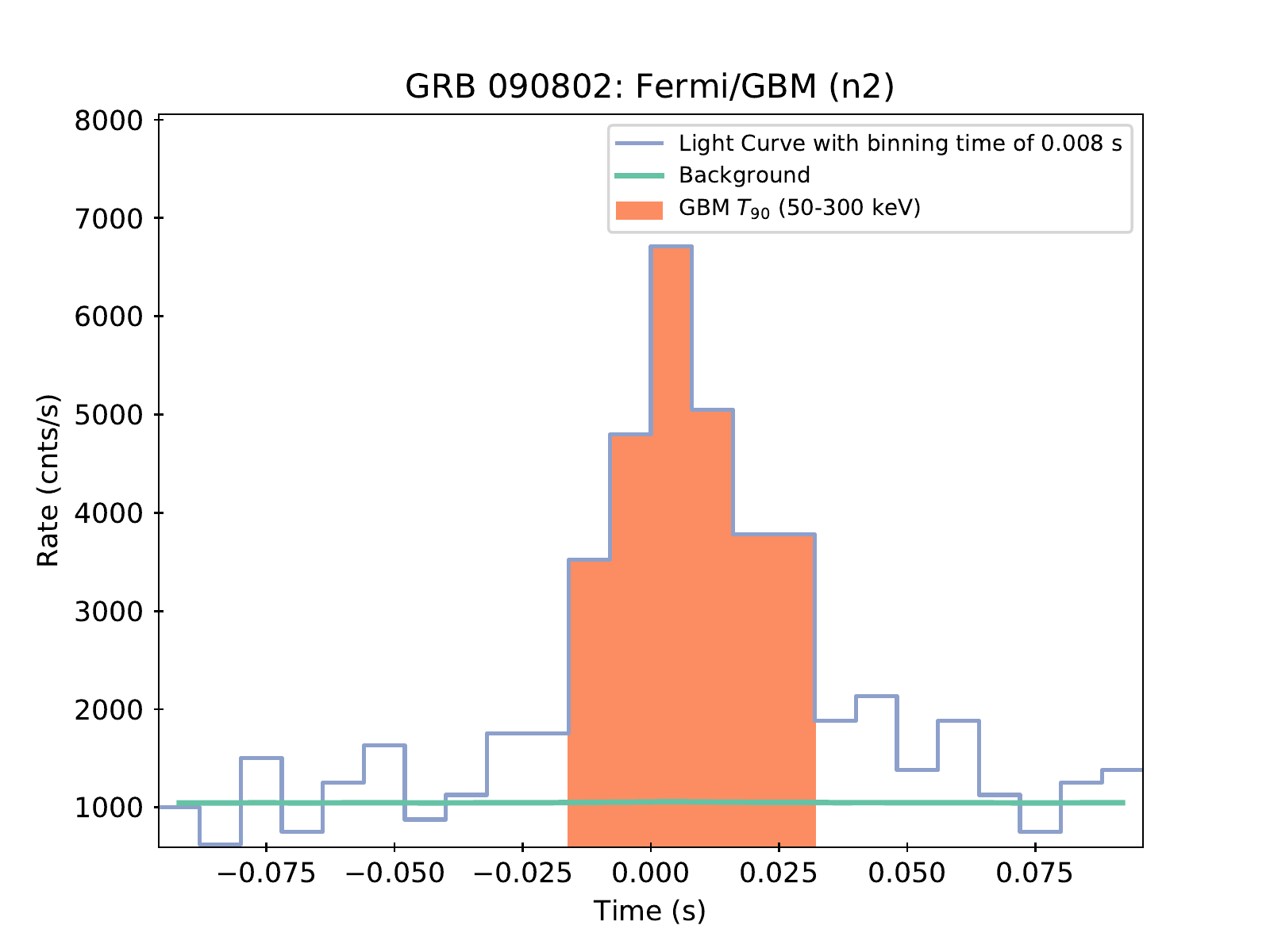}}
		\subfigure[~160822]{
			\includegraphics[width=0.56\textwidth]{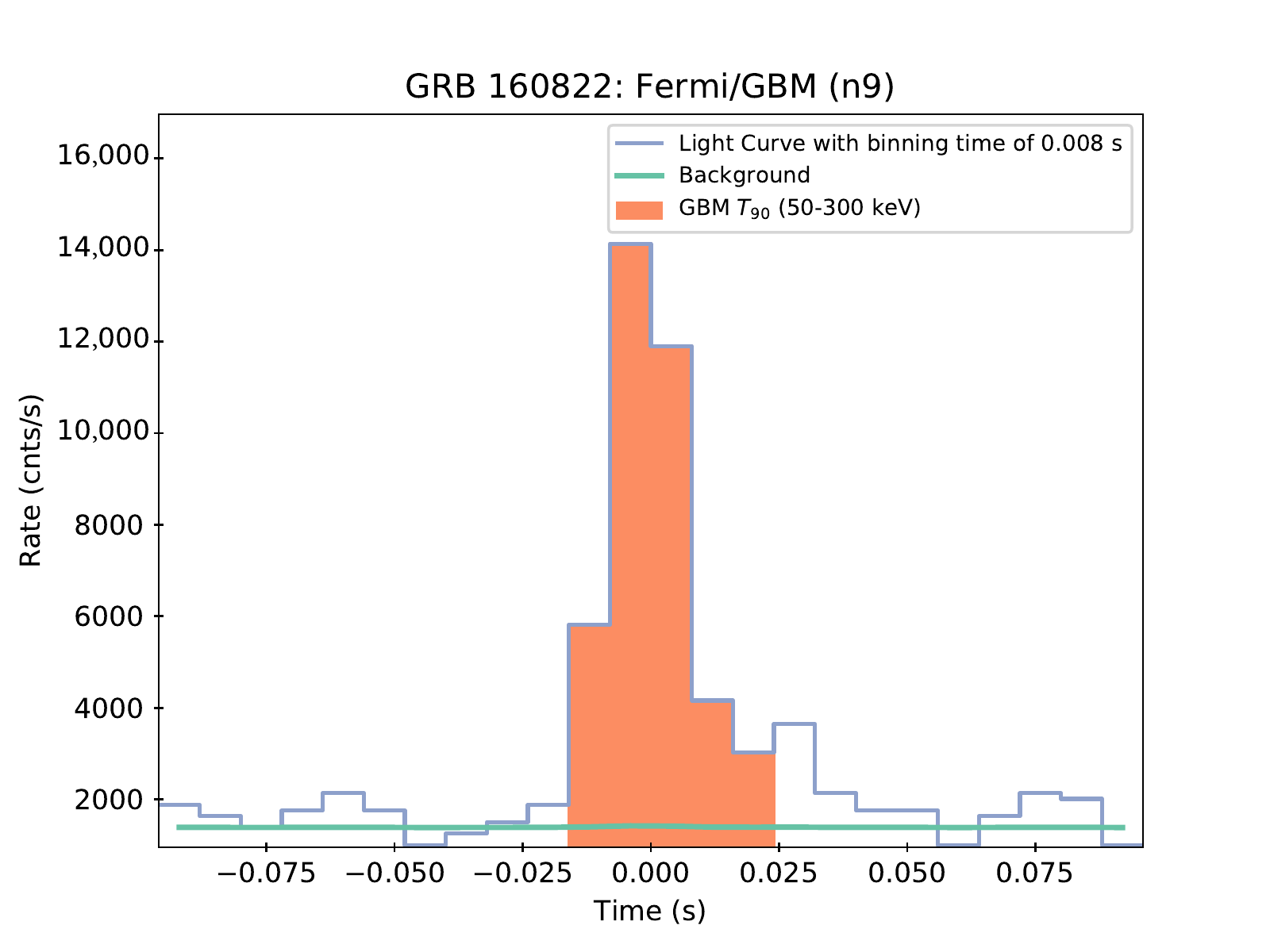}}
\end{adjustwidth}
		\caption{NaI lightcurves of GRB 090802 and GRB 160822. The blue line is the observed lightcurve, the green line is the fitted background line and the orange shadow region is the GBM $T_{90}$ duration. The binned time is 0.008 s.}
		\label{fig:lightcurve}
	\end{figure}

	Fermi/GBM has two types of scintillation detectors, 12  Sodium Iodide units (NaIs, `n0' to `n9', `na' and `nb') and 2 Bismuth Germanate units (BGOs, `b0' and `b1'). NaIs cover the photon energy between about \mbox{8 keV} and \mbox{1 MeV} while BGOs between about \mbox{200 keV} and \mbox{40 MeV}~\citep{Meegan2009}.
	Among 3339 GRBs detected by Fermi-GBM as of August 6, 2022, 14 GRBs are selected with $T_{90}$ (50--300 keV) less than 0.05 s, four of which are excluded with a low signal-to-noise ratio (SNR $<$ 4, ~\cite{Iyyani2021,Ahlgren2019}), e.g., GRBs with the GBM trigger of bn141102112, bn161115745, bn200423579 and bn210119121. Please note that $T_{90}$ is a temporal duration between GBM $T_{05}$ and $T_{95}$, which are the moments when 5$\%$ and 95$\%$ of the total GRB energy fluence is accumulated, respectively. For each GRB, we selected all the detectors as in the Fermi/GBM catalog, as shown in Table ~\ref{tab:info}. Please note that we  abandoned one or more NaI detectors in some GRBs for a low SNR, e.g.,  ‘n1’ in GRB 081229, ‘n8’ + ’n9’ in GRB 091126, and ‘n6’ in GRB 120616. GBM data can be downloaded from the public data site of Fermi/GBM (20 August 2022)~\endnote{\url{https://fermi.gsfc.nasa.gov/ssc/data/access/}.}.
	NaI lightcurves of two GRB are plotted with the binned time of 0.008 s, GRB 090802 and GRB 160822, as shown  in Figure~\ref{fig:lightcurve}.
	
	\begin{table}[H]
		\caption{Information of our GRB sample.\label{tab:info}}
		
\begin{adjustwidth}{-\extralength}{0cm}
\setlength{\tabcolsep}{3.5mm}

	\begin{tabular}{ccccccc}
			\toprule
			\multirow{2}{*}{\textbf{GRB Name}} &\multirow{2}{*}{\textbf{GBM Trigger}} & \boldmath{$T_{90}$}  &\boldmath{ $T_{05}$ } & \boldmath{$T_{95}$}  	 &\multirow{2}{*}{\boldmath{ \textbf{Detetor}}} &\textbf{bkg. Selection} \\
			&   &   \textbf{(s)}&   \textbf{(s)}	&\textbf{(s)}   &  &\textbf{(s)}\\
			\midrule
			081229	&	bn081229675	&	0.032 	&	$-$0.016	&	0.016	&	'n2','n5','b0'	&\ \ \ \ \  [$-$25, $-$10], [15, 30] \\
			090802	&	bn090802235	&	0.048 	&	$-$0.016	&	0.032	&	'n2','n5','b0'	&\ \ \ \ \   [$-$25, $-$10], [15, 30] \\
			091126	&	bn091126389	&	0.024 	&	$-$0.008	&	0.016	&	'n6','n7','nb','b1'	&\ \ \ \ \   [$-$25, $-$10], [15, 30] \\
			120616	&	bn120616630	&	0.048 	&	$-$0.048	&	0	&	'n3','n7','b0'	&\ \ \ \ \   [$-$25, $-$10], [15, 30] \\
			160822	&	bn160822672	&	0.040 	&	$-$0.016	&	0.024	&	'n9','na','b1'	&\ \ \ \ \   [$-$25, $-$10], [15, 30] \\
			171108	&	bn171108656	&	0.032 	&	$-$0.016	&	0.016	&	'n9','nb','b1'	&\ \ \ \ \   [$-$25, $-$10], [15, 30] \\
			180103	&	bn180103090	&	0.016 	&	$-$0.016	&	0	&	'n4','b0','b1'	&\ \ \ \ \   [$-$25, $-$10], [15, 30] \\
			180602	&	bn180602938	&	0.008 	&	$-$0.016	&	$-$0.008	&	'n0','n1','n3','b0'	&\ \ \ \ \   [$-$25, $-$10], [15, 30] \\
			190505	&	bn190505051	&	0.032 	&	$-$0.016	&	0.016	&	'n0','n1','n3','n5','b0'	&\ \ \ \ \   [$-$25, $-$10], [15, 30] \\
			201221	&	bn201221591	&	0.032 	&	$-$0.016	&	0.016	&	'n6','n7','n9','b1'	&\ \ \ \ \   [$-$25, $-$10], [15, 30] \\
			\bottomrule
		\end{tabular}
\end{adjustwidth}
	\end{table}\vspace{-12pt}
	
	\subsection{Spectral Fitting}
	GBM time-tagged event (TTE) data are used in the spectral analysis. Instrument response files are selected with $rsp2$ files. We fit the background rates with auto-selected orders polynomials using the Nelder-Mead method for ten GRBs, two of which are plotted in Figure~\ref{fig:lightcurve}. For NaI detectors, we select photons in channels between about 8 keV and \mbox{1000 keV} and exclude photons in channels between about 33 and 36 keV (Iodine K-edge, ~\citep{Meegan2009}). For BGO detectors, photons in channels between about 200 keV to 40 MeV are included. 
	
	We first select two non-thermal spectral models to fit the gamma-ray spectra, e.g., the Band function model (BAND), the cutoff power-law function model (CPL). Two standard blackbody function (BB)-joined models are employed, the BAND + BB model and the CPL + BB model. For the above four models, BAND model is written as the so-called Band function~\citep{Band1993}, such as
	\begin{eqnarray}
		N_{E}(\rm BAND) = A_{\rm BAND} \left\{ \begin{array}{ll}
			(\frac{E}{E_{\rm piv}})^{\alpha} e^{[-E/E_{0}]}, & E\leq (\alpha-\beta)E_{0} \\
			\\
			(\frac{(\alpha-\beta)E_{0}}{E_{\rm piv}})^{(\alpha-\beta)} e^{(\beta-\alpha)}(\frac{E}{E_{\rm piv}})^{\beta}, & E\geq (\alpha-\beta)E_{0} \\
		\end{array} \right.
		\label{Eq:Band}
	\end{eqnarray}
	where $\alpha$, $\beta$ is the photon index before and after the typical energy of $(\alpha- \beta)E_0$, and ${E_0}$ is the break energy in the $F_E (= E N_E)$ spectrum, and $E_{\rm peak}= (2+\alpha) E_0$ is the peak energy in the $E F_E (= E^2 N_E)$ spectrum~\citep{Norris2005}. CPL could be regarded as the lower energy segment of the BAND model but with an exponential cutoff-power-law decay in the high-energy band, such as
	\begin{equation}
		N_{E}({\rm CPL}) = A_{\rm CPL} (\frac{E}{E_{\rm piv}})^{\Gamma} e^{-E/E_{\rm c}},
	\end{equation}
	where $\Gamma$ is the photon index and $E_{\rm c}$ is the cutoff energy. $E_{\rm piv}$ in both models is the pivot energy and fixed at 100 keV.
	BB component is usually modified by the standard Planck spectrum, which is given by the photon flux,
	\begin{equation}
		N_{E}({\rm BB}) =  A_{\rm BB}  \frac{E^2}{exp[E/kT]-1},
		\label{Eq:BB}
	\end{equation}
	where $k$ is the Boltzmann's constant, and the joint parameter $kT$ as an output parameter in common. In all spectral models, $A$ is the amplitude factor.
	
	We perform the spectral fitting in the Multi-Mission Maximum-Likelihood package~(3ML; \citep{Vianello2015}), which provides a common high-level interface and model definition. We create the spectral files in the GBM $T_{90}$ duration, and use the maximum-likelihood-estimation (MLE) method to fit the data with above four spectral models. In the 3ML package, \textls[-20]{the command named $JointLikelihood$ is employed to carry out the spectral fit, which returns the values of the maximum-likelihood values($L$), the spectral parameters and the photon fluxes. 
	The package also returns the values of the Annika Information Criterion (\mbox{AIC = $-$2ln$L$+2$n_{para}$)} and the Bayesian Information Criterion (\mbox{BIC = $-$2ln$L$+$n_{para} \ln n_{data}$})~\mbox{\citep{Akaike1974,Schwarz1978}},} where $n_{para}$ is the number of spectral parameters and $n_{data}$ is the total number of the observational \mbox{data points.} 
	
	\section{Results \label{results}}
	
	\subsection{General Results}
	We obtained the best-fit spectral parameters, the likelihood values ($-$ln($L$)), the degree of freedom (DoF), the AIC values and the BIC values in four spectral models. For the\mbox{ BAND + BB} and \mbox{CPL + BB} models, the flux ratio ($R$) between the BAND (or CPL) component and the BB component is also given. We sorted the models by the BIC values, which can be found in Table~\ref{tab:sed}. Here, C, B, C2 and B2 represent the models of the CPL, the BAND, the CPL + BB and the BAND +BB, respectively. We discussed the results of each model following. As seen in Table~\ref{tab:sed}, the model sequence according to the order of small BIC to large BIC is C, B, C2 and B2 except for two GRBs (GRB 091126 and GRB 180602); therefore we presented the results from the CPL model, the BAND model, the CPL + BB model to the BAND + BB model.
	
	\begin{table}[H]
		\caption{Derived spectral parameters of four spectral models for each GRB in our sample.\label{tab:sed}}
		\tabcolsep=0.1cm
	
\begin{adjustwidth}{-\extralength}{0cm}
	\setlength{\tabcolsep}{1.95mm}
\begin{tabular}{ccccccccc}
			\toprule
			\multirow{2}{*}{\boldmath{$\rm GRB$}}&\multirow{2}{*}{\boldmath{$\rm M~^1$}}&\multirow{2}{*}{\textbf{\boldmath{AIC/BIC/$-$ln($L$)}}}&\multirow{2}{*}{\textbf{\boldmath{DoF~$^{\rm 2}$}}}&\multirow{2}{*}{\boldmath{$\alpha / \Gamma$}}&\boldmath{$E_{\rm peak}$}&\multirow{2}{*}{\boldmath{$\beta$}}&\boldmath{${kT}$}&\multirow{2}{*}{\boldmath{${R~^{\rm 3}}$}}\\
			&&&&& \textbf{(keV) }&& \textbf{(keV)}&\\
			\midrule
			081229&C&$-$1741.1/$-$1729.4/$-$873.5 &356&$-$0.04 $\pm$ 0.43 &358 $\pm$ 184 & - & - & -\\
			…&B&$-$1739.3/$-$1723.8/$-$873.7 &355&0.21 $\pm$ 0.52 &290 $\pm$ 82 &$-$2.64 $\pm$ 0.80 & - & -\\
			…&C2&$-$1738.4/$-$1719.0/$-$874.2 &354&0.50 $\pm$ 1.46 &443 $\pm$ 312 & - &26 $\pm$ 10 &0.10\\
			…&B2&$-$1736.4/$-$1713.1/$-$874.2 &353&0.51  $\pm$  1.12 &443 $\pm$ 163 &$-$8.49 $\pm$ 2.87 &26 $\pm$ 10&0.11 \\
			\midrule
			090802&C&$-$1151.8/$-$1140.1/$-$578.9 &356&$-$0.42 $\pm$ 0.14 &376 $\pm$ 82 & - & - & -\\
			…&B&$-$1152.3/$-$1136.7/$-$580.1 &355&$-$0.30 $\pm$ 0.18 &310 $\pm$ 55 &$-$2.28 $\pm$ 0.29 & - & -\\
			…&C2&$-$1148.0/$-$1128.5/$-$579.0 &354&$-$0.49 $\pm$ 0.20 &418 $\pm$ 156 & - &43 $\pm$ 27&0.04 \\
			…&B2&$-$1148.3/$-$1125.0/$-$580.1 &353&$-$0.30 $\pm$ 0.11 &310 $\pm$ 1 &$-$2.28 $\pm$ 0.27 &54 $\pm$ 1&$<$0.01 \\
			\midrule
			091126&C&$-$2907.7/$-$2895.2/$-$1456.8 &476&$-$0.09 $\pm$ 0.45 &711 $\pm$ 385 & - & - & -\\
			…&C2&$-$2910.6/$-$2889.7/$-$1460.3 &474&0.95 $\pm$ 0.92 &536 $\pm$ 383 & - &7 $\pm$ 1&0.02 \\
			…&B&$-$2905.7/$-$2889.0/$-$1456.8 &475&$-$0.09 $\pm$ 0.41 &711 $\pm$ 192 &$-$9.98 $\pm$ 4.07 & - & -\\
			…&B2&$-$2908.6/$-$2883.6/$-$1460.3 &473&0.95 $\pm$ 0.89 &536 $\pm$ 133 &$-$10.00 $\pm$ 2.12 &7 $\pm$ 2&0.02 \\
			\midrule
			120616&C&$-$895.7/$-$884.0/$-$450.8 &358&$-$0.11 $\pm$ 0.50 &744 $\pm$ 549 & - & - & -\\
			…&B&$-$893.7/$-$878.1/$-$450.8 &357&$-$0.11 $\pm$ 0.44 &746 $\pm$ 290 &$-$9.94 $\pm$ 4.20 & - & -\\
			…&C2&$-$891.9/$-$872.5/$-$451.0 &356&3.33 $\pm$ 1.51 &608 $\pm$ 221 & - &40 $\pm$ 12&0.14 \\
			…&B2&$-$889.7/$-$866.4/$-$450.9 &355&$-$0.04 $\pm$ 0.48 &721 $\pm$ 237 &$-$10.00 $\pm$ 0.05 &4 $\pm$ 3&$<$0.01 \\
			\midrule
			160822&C&$-$1185.8/$-$1174.2/$-$595.9 &355&$-$0.59 $\pm$ 0.11 &590 $\pm$ 139 & - & - & -\\
			…&B&$-$1187.3/$-$1171.8/$-$597.6 &354&$-$0.57 $\pm$ 0.11 &549 $\pm$ 98 &$-$2.36 $\pm$ 0.33 & - & -\\
			…&C2&$-$1187.9/$-$1168.5/$-$598.9 &353&0.02 $\pm$ 1.25 &587 $\pm$ 426 & - &19 $\pm$ 3&0.12 \\
			…&B2&$-$1189.3/$-$1166.0/$-$600.6 &352&0.02 $\pm$ 0.73 &559 $\pm$ 124 &$-$2.44 $\pm$ 0.36 &19 $\pm$ 3&0.05 \\
			\midrule
			171108&C&$-$1724.0/$-$1712.4/$-$865.0 &356&$-$0.25 $\pm$ 0.22 &106 $\pm$ 20 & - & - & -\\
			…&B&$-$1722.0/$-$1706.5/$-$865.0 &355&$-$0.25 $\pm$ 0.17 &106 $\pm$ 1 &$-$9.00 $\pm$ 2.28 & - & -\\
			…&C2&$-$1722.6/$-$1703.2/$-$866.3 &354&1.18 $\pm$ 1.17 &126 $\pm$ 41 & - &10 $\pm$ 2&0.24 \\
			…&B2&$-$1722.8/$-$1699.5/$-$867.4 &353&5.00 $\pm$ 0.01 &122 $\pm$ 16 &$-$3.78 $\pm$ 0.76 &11 $\pm$ 1&0.62 \\
			\midrule
			180103&C&$-$2498.2/$-$2486.6/$-$1252.1 &359&$-$0.33 $\pm$ 0.43 &819 $\pm$ 519 & - & - & -\\
			…&B&$-$2498.2/$-$2482.7/$-$1253.1 &358&0.09 $\pm$ 0.63 &527 $\pm$ 188 &$-$2.18 $\pm$ 0.30 & - & -\\
			…&C2&$-$2495.5/$-$2476.0/$-$1252.8 &357&$-$0.67 $\pm$ 0.56 &1847 $\pm$ 1861 & - &90 $\pm$ 34&0.16 \\
			…&B2&$-$2495.0/$-$2471.7/$-$1253.5 &356&0.24 $\pm$ 0.58 &504 $\pm$ 138 &$-$2.18 $\pm$ 0.30 &2 $\pm$ 1&$<$0.01 \\
			\midrule
			180602&C&$-$3928.5/$-$3916.0/$-$1967.3 &474&0.16 $\pm$ 0.27 &420 $\pm$ 104 & - & - & -\\
			…&C2&$-$3924.5/$-$3903.7/$-$1967.3 &472&0.16 $\pm$ 0.27 &419 $\pm$ 101 & - &47 $\pm$ 1 &$<$0.01\\
			…&B2&$-$3924.5/$-$3899.5/$-$1968.2 &471&$-$0.51 $\pm$ 0.57 &315 $\pm$ 267 &$-$1.94 $\pm$ 0.45 &98 $\pm$ 14&0.85 \\
			…&B&$-$3806.5/$-$3789.8/$-$1907.2 &473&$-$1.50 $\pm$ 0.27 &528 $\pm$ 82433 &$-$1.50 $\pm$ 0.00 & - & -\\
			\midrule
			190505&C&$-$2856.8/$-$2843.6/$-$1431.4 &592&$-$0.30 $\pm$ 0.20 &309 $\pm$ 80 & - & - & -\\
			…&B&$-$2854.8/$-$2837.2/$-$1431.4 &591&$-$0.30 $\pm$ 0.19 &309 $\pm$ 44 &$-$7.39 $\pm$ 2.83 & - & -\\
			…&C2&$-$2852.8/$-$2830.9/$-$1431.4 &590&$-$0.38 $\pm$ 0.19 &327 $\pm$ 84 & - &50 $\pm$ 30&0.04 \\
			…&B2&$-$2850.8/$-$2824.4/$-$1431.4 &589&$-$0.35 $\pm$ 0.14 &322 $\pm$ 1 &$-$9.99 $\pm$ 0.39 &43 $\pm$ 1&0.12 \\
			\midrule
			201221&C&$-$2133.1/$-$2120.6/$-$1069.5 &476&$-$0.63 $\pm$ 0.14 &371 $\pm$ 105 & - & - & -\\
			…&B&$-$2131.2/$-$2114.5/$-$1069.6 &475&$-$0.57 $\pm$ 0.17 &333 $\pm$ 72 &$-$2.95 $\pm$ 0.94 & - & -\\
			…&C2&$-$2132.9/$-$2112.1/$-$1071.5 &474&$-$0.83 $\pm$ 0.19 &621 $\pm$ 352 & - &33 $\pm$ 7&0.10 \\
			…&B2&$-$2130.9/$-$2105.9/$-$1071.5 &473&$-$0.83 $\pm$ 0.17 &622 $\pm$ 227 &$-$9.53 $\pm$ 3.83 &33 $\pm$ 7&0.11 \\
			\bottomrule
		\end{tabular}
\end{adjustwidth}
	\noindent\footnotesize{$^{1}$ B: BAND model; C: CPL model; C2: CPL + BB model; B2: BAND + BB model. $^{2}$ DoF: Degree of freedom.
		$^{3}$ $R$: Ratio between the BB flux and BAND/CPL flux (8 keV--40 MeV).}	
	\end{table}
	
	\subsubsection{Result for the CPL model}
	For the CPL model, the photon index $\Gamma$ is between about $-$0.63 and 0.16 for all GRBs in our sample, all of which show a hard spectral index when being compared with the typical BAND low-energy photon index of $-$1.0. The peak energy $E_{peak}$ is between about 106 keV and 819 keV in our sample, all of which share a hard peak energy (the typical BAND peak energy is about 250 keV) except for GRB 171108 with $E_{peak}$ of 106 $\pm$ 20 keV. 
	
\begin{figure}[H]

\begin{adjustwidth}{-\extralength}{0cm}
\centering 
		\includegraphics[width=0.6\textwidth]{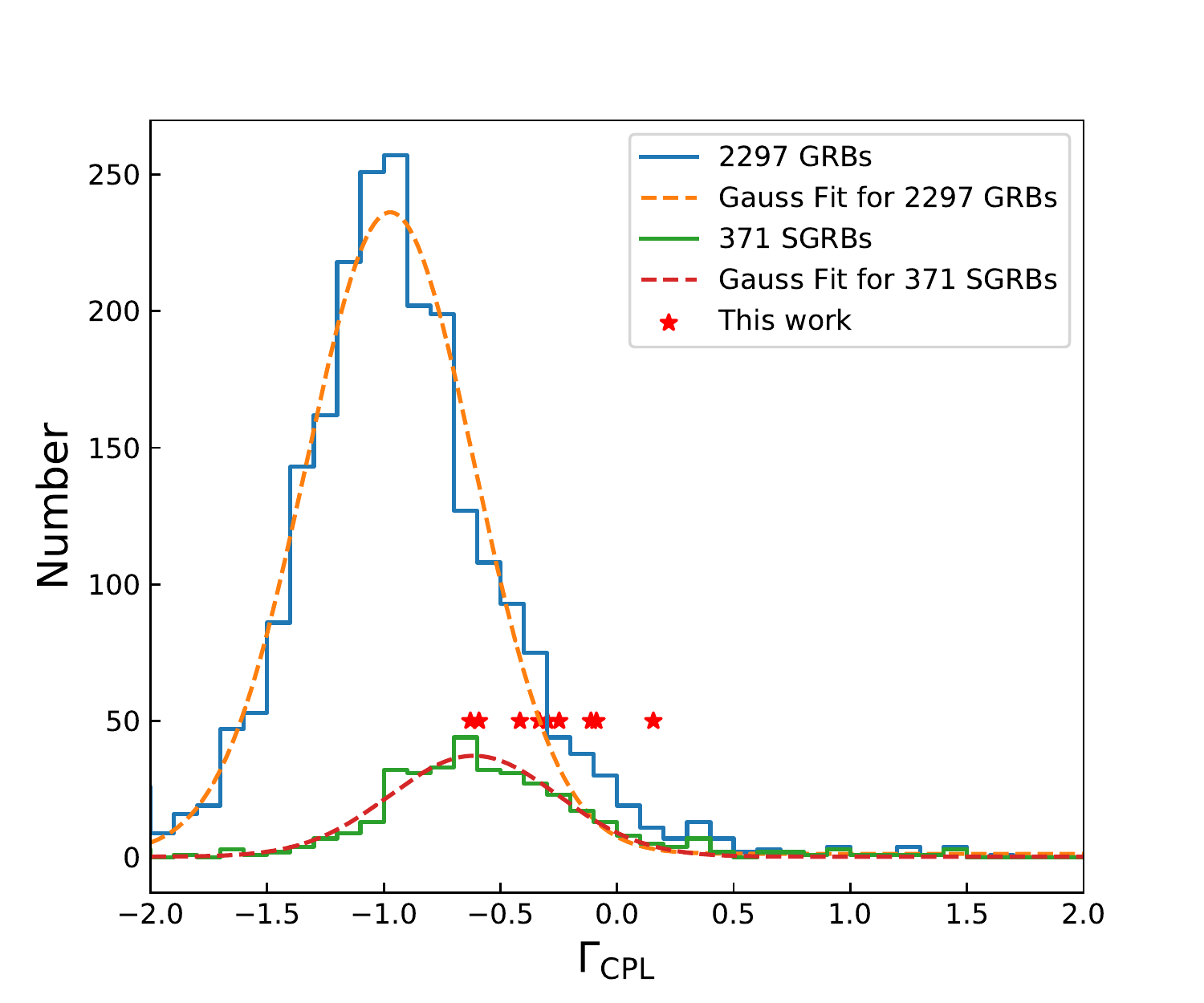}
		\includegraphics[width=0.6\textwidth]{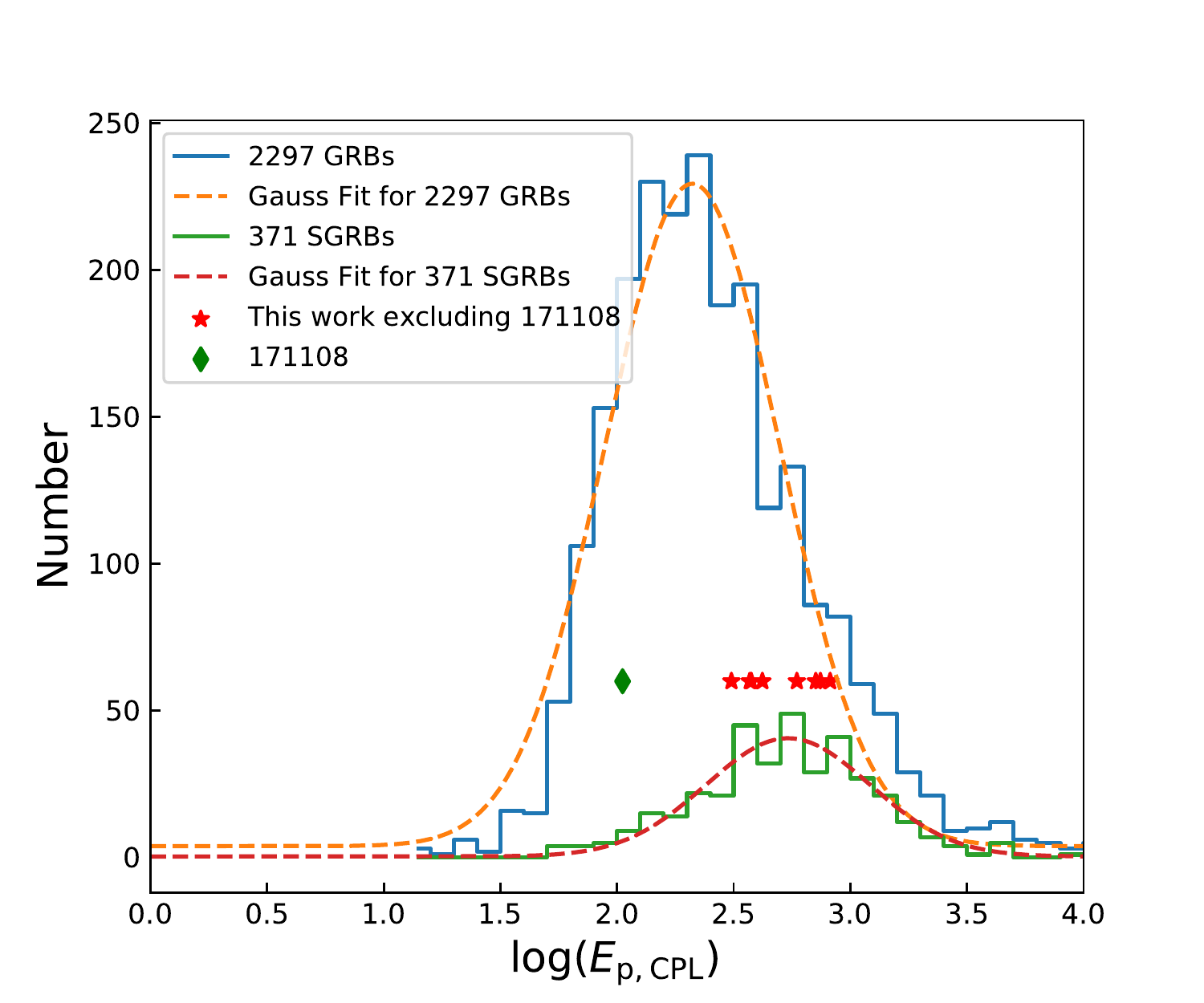}
\end{adjustwidth}
		\caption{Comparisions of the CPL spectral parameters between three samples: the catalog sample, the SGRB sample and the sample in this work. (\textbf{Left}) $\Gamma_{\rm cata} = -0.97\pm0.72$, $\Gamma_{\rm SGRB} = -0.61\pm0.72$. (\textbf{Right}) $E_{\rm peak,\ cata} = 10^{2.32~\pm~0.75}$ keV, $E_{\rm peak,\ SGRB} = 10^{2.73~\pm~0.70}$ keV.}
		\label{fig:compare1}
	\end{figure}

	Figure~\ref{fig:compare1} shows the comparison of the CPL spectral parameters between our sample and the Fermi/GBM-catalog sample~\cite{vonKienlin2020}. 
	First, as of August 6, 2022, the time-integrated spectra of 2297 GRBs (hereafter catalog sample) are fitted by the CPL model, the photon index ($\Gamma_{\rm cata}$) is between about $-$2.5 and 10.7, whose distribution can be modeled by a Gaussian function $y=f(x)$ as
	\begin{equation}
		y =  y_0 + \frac{C}{\sigma\sqrt{2\pi}}e^{-2\frac{(x-\mu)^2}{\sigma^2}},
		\label{Eq:Gauss}
	\end{equation}
	where $y_0,C$ are the free parameters and $\sigma,\mu$ are the expect value  and the standard deviation (SD), respectively. The Gaussian fitting results in an expected index value $\Gamma_{\rm cata}$ of $-0.97$ with SD of 0.72. We also plotted a sub-sample of 371 SGRBs (hereafter SGRB sample), whose Gaussian Fitting results in an expected index value $\Gamma_{\rm SGRB}$ of $-0.61$ with SD of 0.72. We found that the photon index ($\Gamma_{\rm this}$) in our sample is harder than the expect photon index in two other sample, which can be found in the left panel of Figure~\ref{fig:compare1}. 
	
	Secondly, the distributions of the peak energy in these three samples are plotted in the right panel of Figure~\ref{fig:compare1}. Except for GRB 171108, we found that the peak energy ($E_{\rm peak,\ this}$) in our sample is harder than the expect peak energy in the catalog sample, e.g., \mbox{$E_{\rm peak,\ cata} = 210.4$ keV}. Most of GRBs in our sample have a hard peak energy compared to that of the SGRB sample, e.g., $E_{\rm peak,\ SGRB} = 540.8$ keV.
	
	In short, except for GRB 171108, the results both of the photon index and the peak energy indicate a common hard spectral component in our sample, which is strongly in favor of the typical GRB classification, such as the short/hard GRB and the long/soft GRB~\citep{Kouveliotou1993,Zhang2009,Burns2018}. 
	
	\vspace{-4pt}
	
	\subsubsection{Result for the BAND Model}
	For the BAND model, excluding GRB 180602, the low-energy photon index $\alpha_{\rm this}$ is between about $-$0.57 and 0.21, which shows a hard photon index. Please note that for GRB 180602, it was found that the BAND model is not a candidate model to fit its spectrum because of the similar low-energy index and high-energy index ($\alpha\simeq\beta$), which results in an unreasonable error of the peak energy of $528\pm82433$ keV. Therefore, we discussed the results excluding GRB 180602 in this section. The high-energy photon index $\beta_{\rm this}$ is between about $-$9.98 and $-$2.18 in our sample. $\beta_{\rm this}$ in four GRBs has a very soft index, e.g., $\beta < $-$7.0$, in which case the CPL model is usually a better model than the BAND model to fit the data. The peak energy ($E_{\rm peak,\ this}$) is between about 106 keV and 746 keV, which implies a hard peak energy except for GRB 171108 with a peak energy of 106 $\pm$ 1 keV. 
	
\begin{figure}[H]

\begin{adjustwidth}{-\extralength}{0cm}
\centering 
		\includegraphics[width=0.58\textwidth]{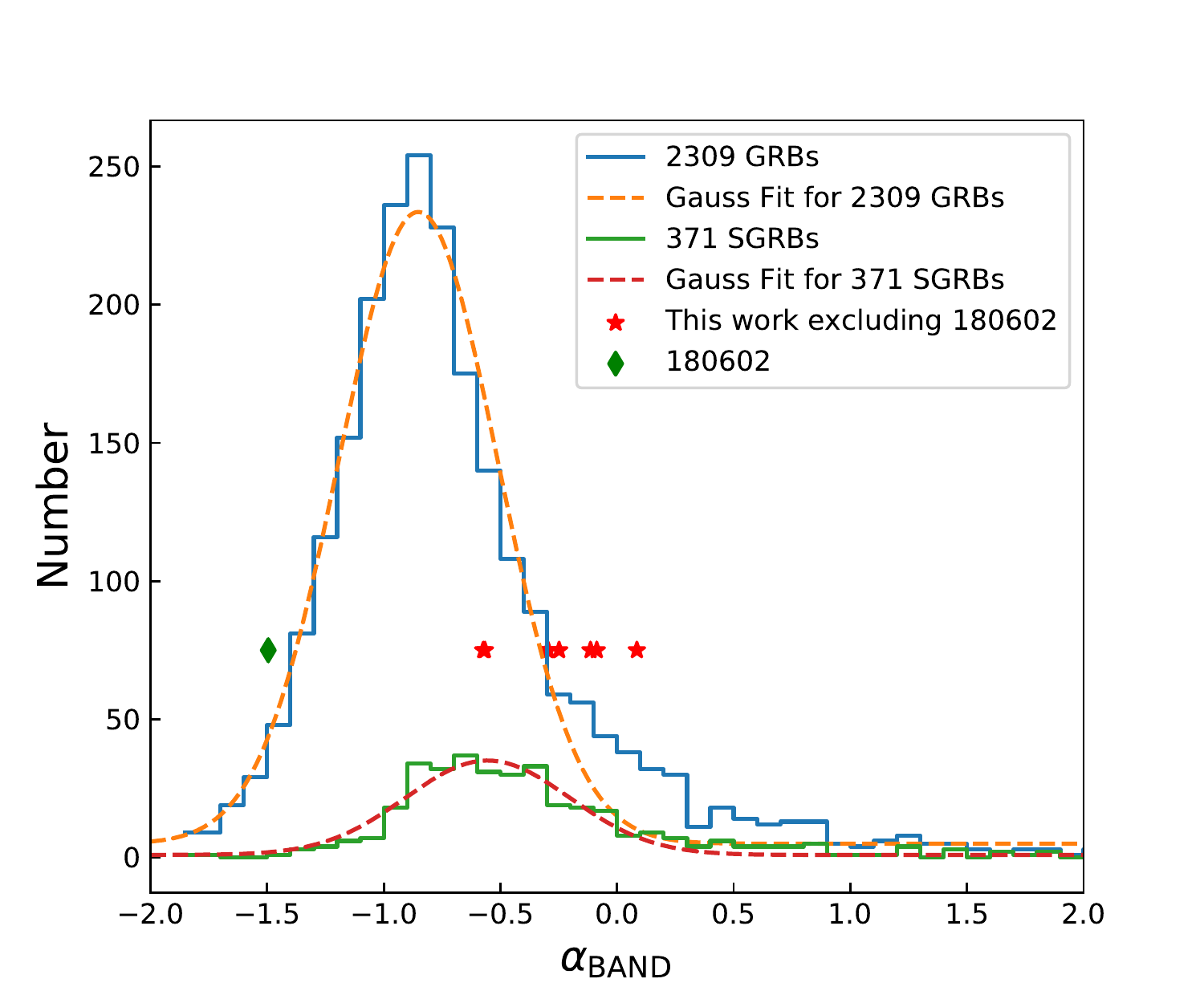}
		\includegraphics[width=0.58\textwidth]{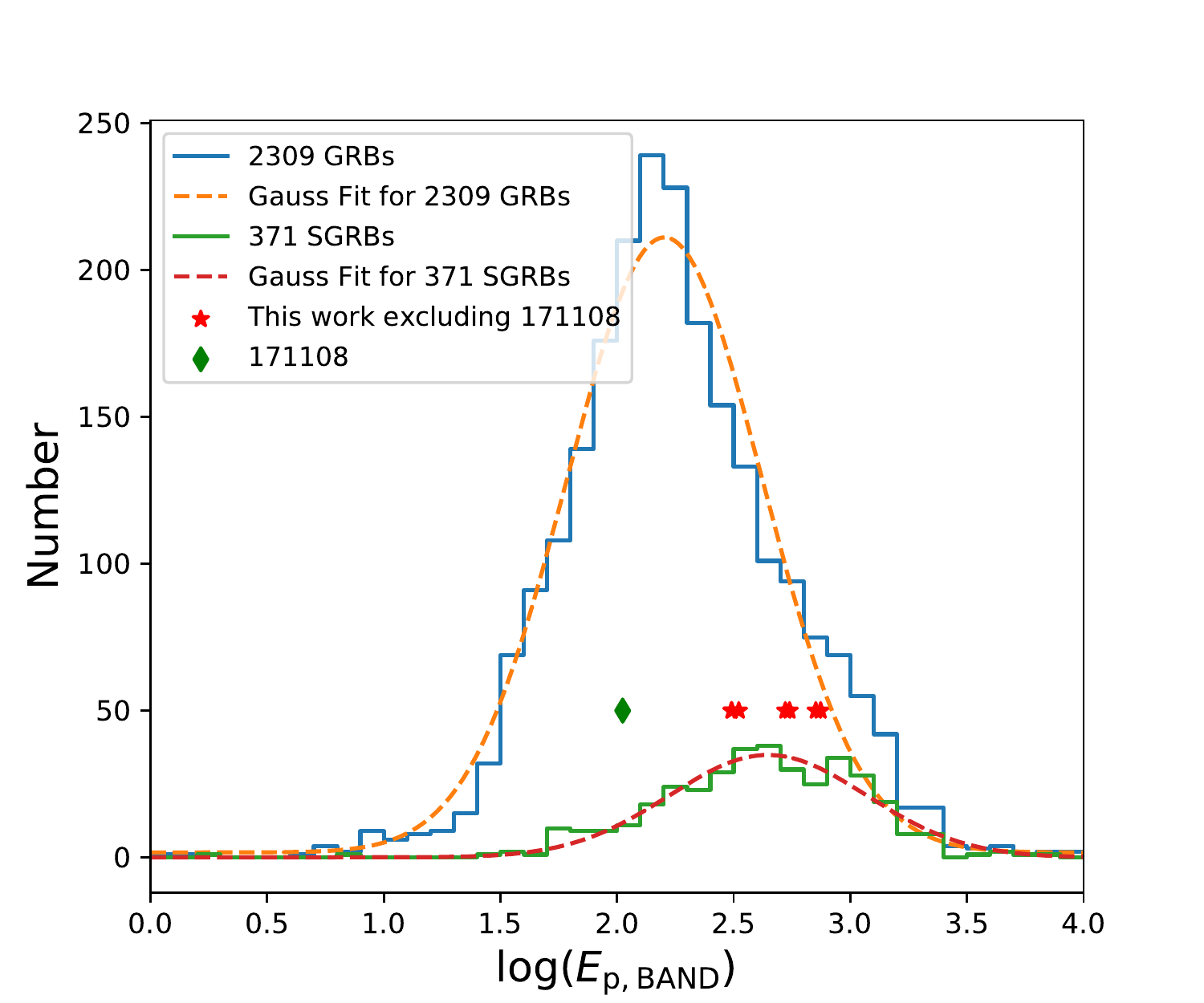}	
		\includegraphics[width=0.58\textwidth]{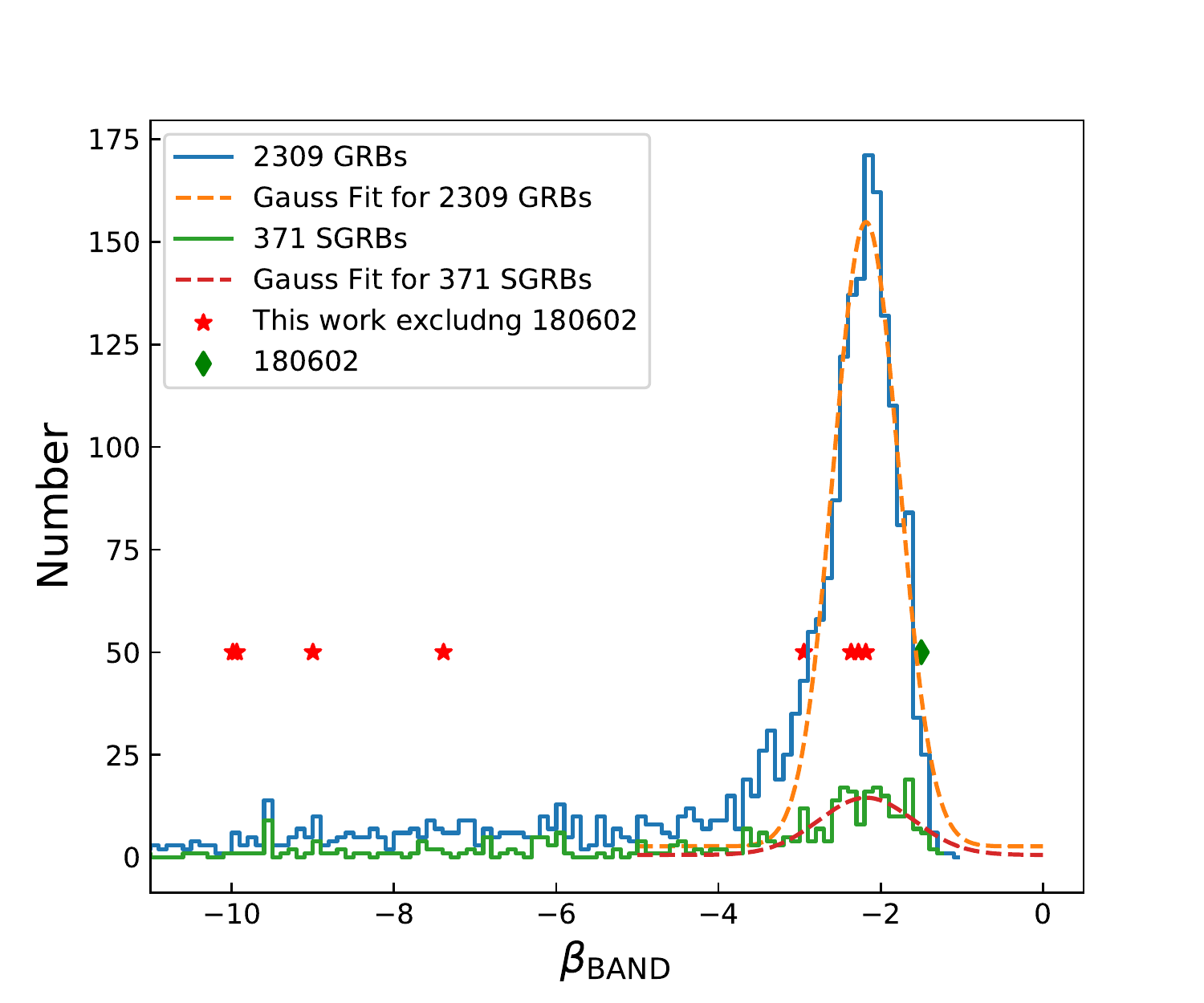}	
\end{adjustwidth}
		\caption{Same as Figure~\ref{fig:compare1}, but for the BAND model. (\textbf{Top Left}) $\alpha_{\rm cata} = -0.85\pm0.68$, \mbox{$\alpha_{\rm SGRB} = -0.56\pm0.70$}. (\textbf{Top Right}) $E_{\rm peak,\ cata} = 10^{2.20\pm0.84}$ keV, $E_{\rm peak,\ SGRB} = 10^{2.65\pm0.84}$ keV. (\textbf{Bottom}) $\beta_{\rm cata} = -2.18\pm0.81$, $\beta_{\rm SGRB} = -2.18\pm1.17$.}
		\label{fig:compare2}
	\end{figure}

	We compared the results with the catalog sample (2309 GRBs) and the SGRB sample (371 GRBs), which can be found in Figure~\ref{fig:compare2}. The same method is performed as above. For the low-energy photon index in the catalog sample and the SGRB sample is $\alpha_{\rm cata} = -0.85$ with SD of 0.68,  $\alpha_{\rm SGRB} = -0.56$ with SD of 0.70, respectively. Almost all of GRBs in our sample have a hard low-energy photon index. For the high-energy photon index in the catalog sample, the SGRB sample is $\beta_{\rm cata} = -2.18$ with SD of 0.81,  $\beta_{\rm SGRB} = -2.18$ with SD of 1.17, respectively. We found that all GRB in our sample have a soft high-energy photon index, e.g., $\beta_{\rm this} < -2.1$. For the BAND peak energy, the Gaussian fitting results in $E_{\rm peak,\ cata} = 159.7$ keV, $E_{\rm peak,\ SGRB} =443.5$ keV for the catalog sample and the SGRB sample, respectively. Except for GRB 171108, it is also found that the BAND peak energy in our sample is larger than that in the catalog sample. However, the BAND peak energy in our sample is around the expect value of the peak energy in the SGRB sample.
	
	As a short conclusion for the BAND model, the hard low-energy photon index in our sample is harder than the theoretical predictive low-energy photon index by a slow-cooling synchrotron ($-$2/3) or a fast-cooling synchrotron ($-$3/2)~\citep{Sari1998,Tang2021}.
	
	\subsubsection{Results for the models with BB component}
	For the CPL + BB model, the peak energy of the CPL component is between about \mbox{126 keV} and 1847 keV, the photon index is between about $-$0.83 and 3.33. The BB temperature is between about 7 keV and 90 keV. 
	
	For the BAND + BB model, the peak energy of the BAND component is between about 122 keV and 721 keV, the low-energy photon index is between about $-$0.83 and 5.00, and the high-energy photon index is between about $-$10.00 and $-$1.500. The BB temperature in this model is between about 2 keV and 98 keV. 
	
	To test whether a thermal component dominate the gamma-ray spectrum in a GRB of our sample, we calculated the ratio ($R$) between the BB flux and the BAND/CPL flux in the energy range from 8 keV to 40 MeV, which is presented in Table~\ref{tab:sed}. $R$ ranges from $< 10^{-4}$ to 0.846, most of which are below 0.1. We thus discuss the results in three classes, such as (i): both \mbox{CPL + BB} and the \mbox{BAND + BB} model with $R > 0.1$; (ii) both \mbox{CPL + BB} and the \mbox{BAND + BB} model with $R < 0.1$; (iii) Only one of these two models \mbox{with $R > 0.1$}.
	
\begin{figure}[H]

\begin{adjustwidth}{-\extralength}{0cm}
\centering 
		\subfigure[~BAND]{
			\includegraphics[width=0.47\textwidth]{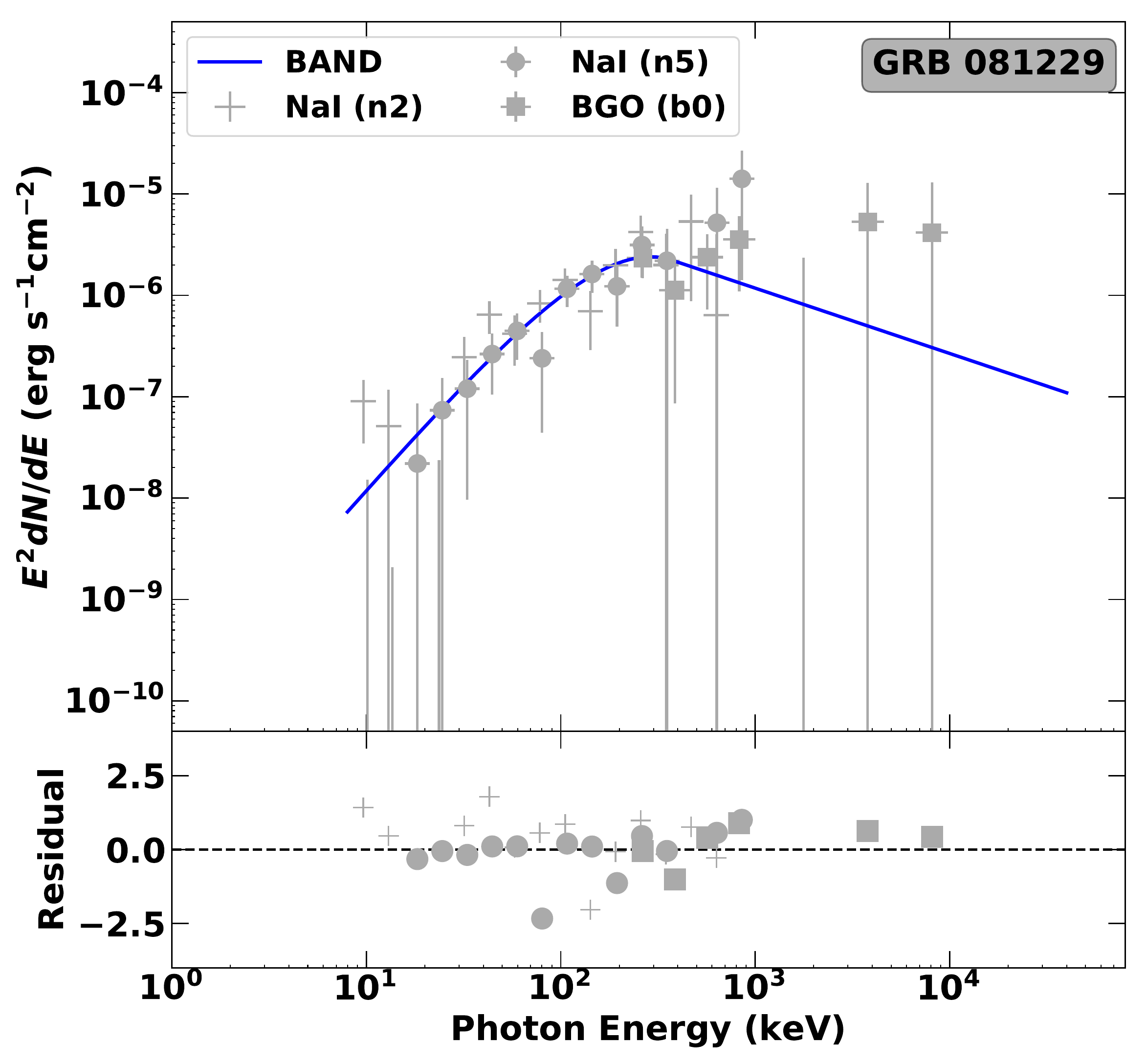}}\quad\quad\quad\quad\quad
		\subfigure[~CPL]{
			\includegraphics[width=0.47\textwidth]{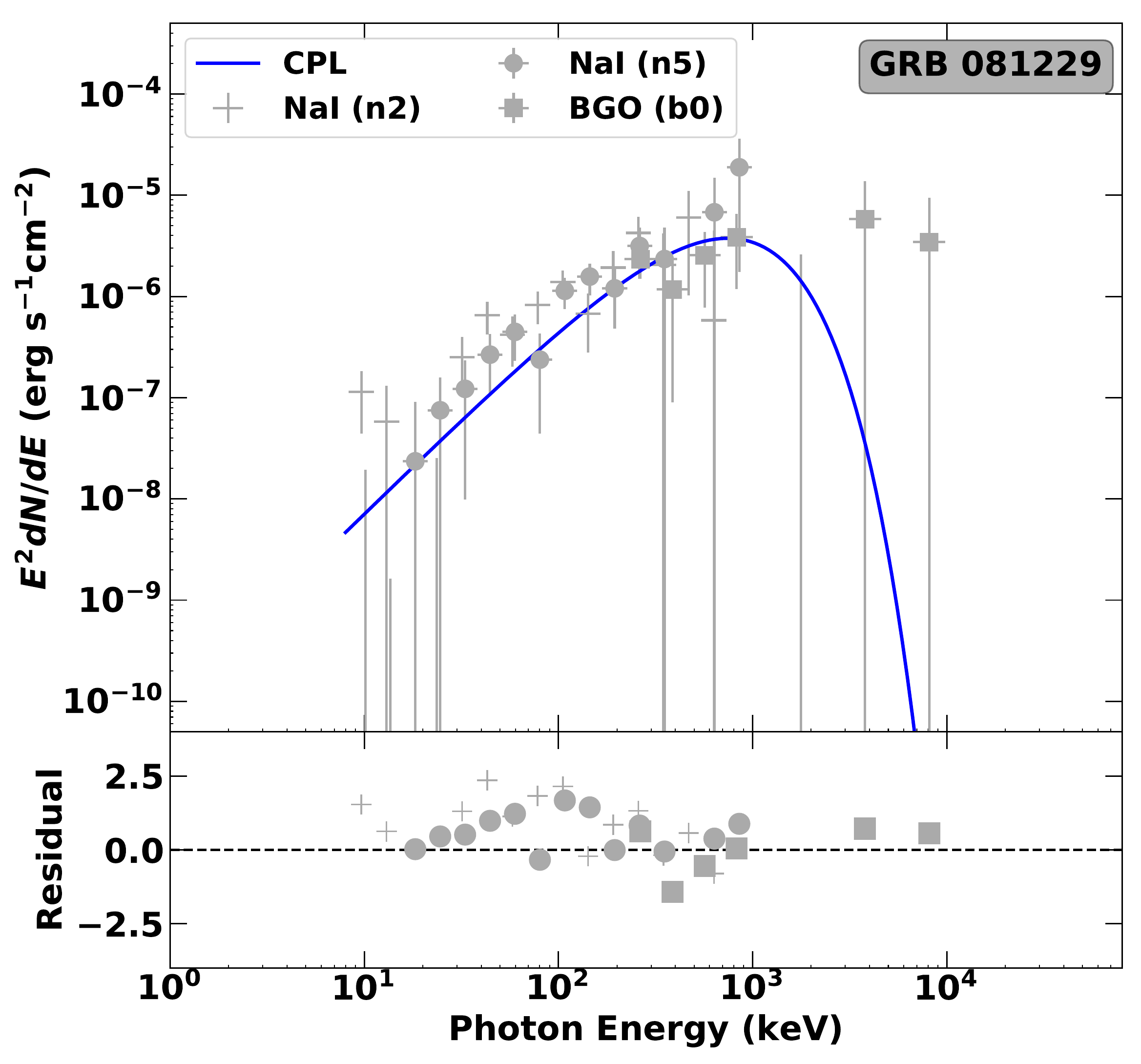}}\\
		\subfigure[~BAND + BB]{
			\includegraphics[width=0.47\textwidth]{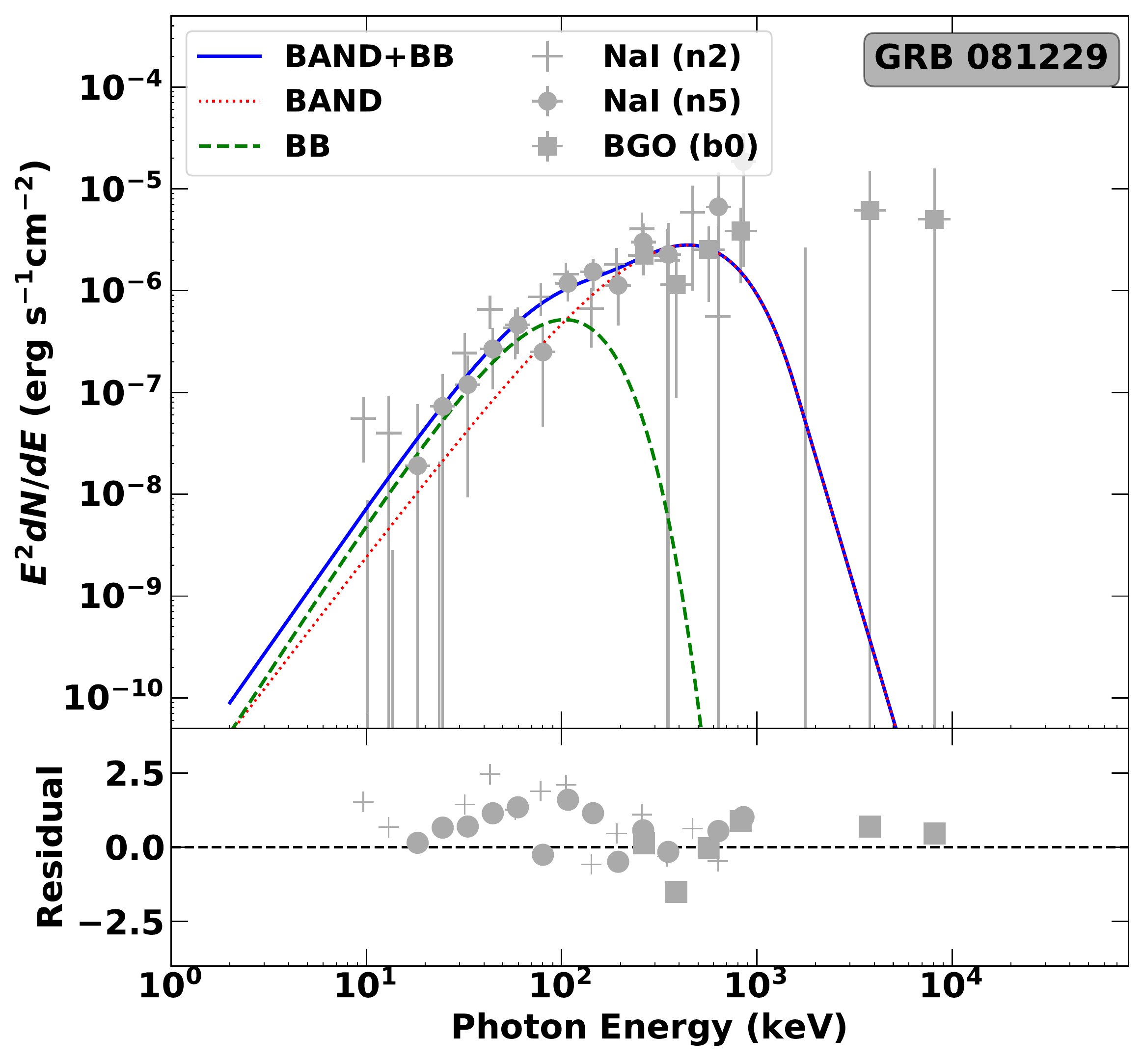}}\quad\quad\quad\quad\quad
		\subfigure[~CPL + BB]{
			\includegraphics[width=0.47\textwidth]{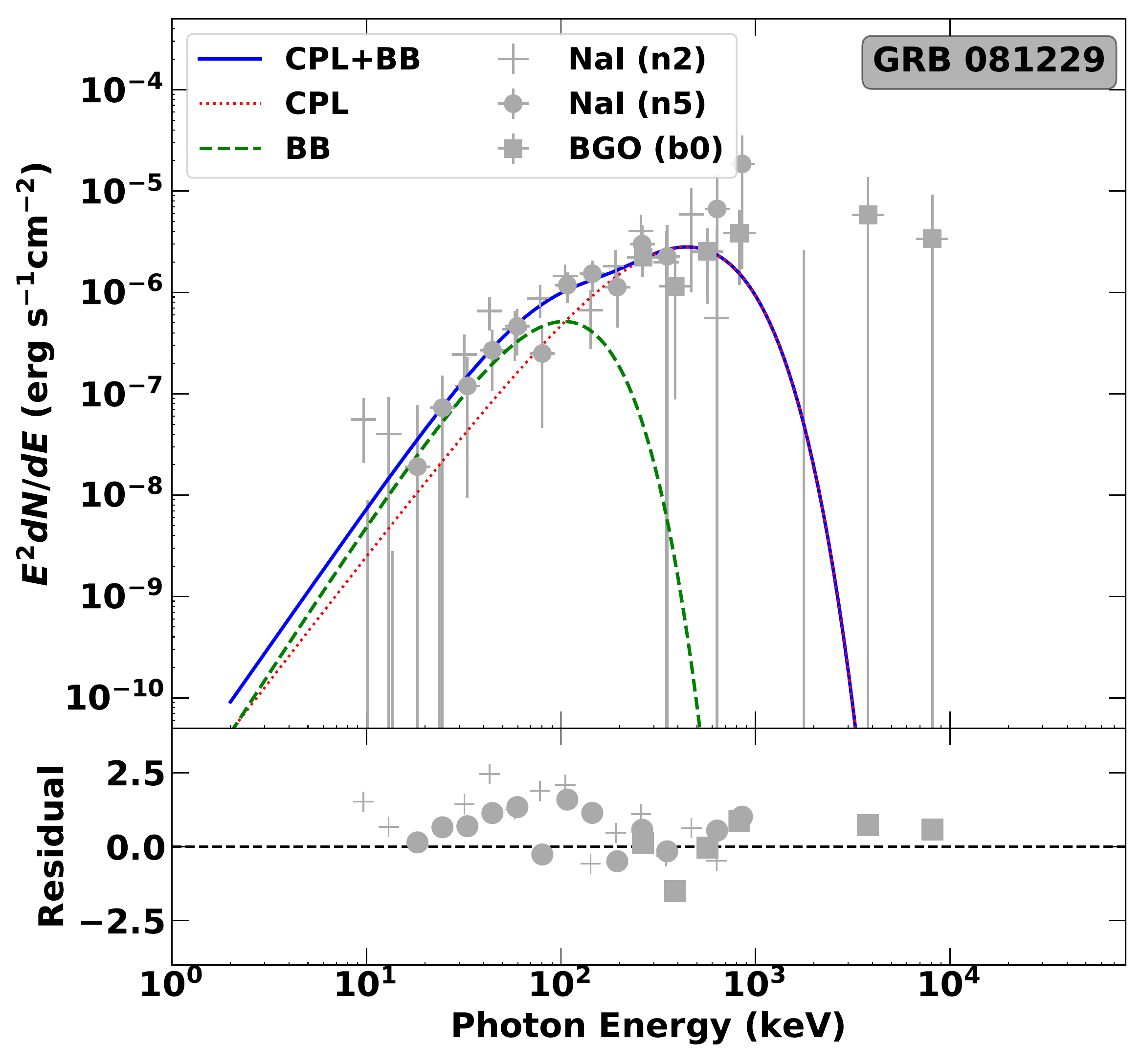}}
\end{adjustwidth}
		\caption{Spectral-fitting results for one GRB in class (i) that with $R > 0.1$ in both of the \mbox{BAND + BB} model and the CPL + BB model, such as GRB 081221.}
		\label{fig:subsample1}
	\end{figure}

	For class (i), three GRBs are included, GRB 081229, GRB 171108 and GRB 201221. The results of GRB 081229 are plotted in Figure~\ref{fig:subsample1}. Thermal component in this class of GRBs usually dominates the observed photon flux at the low-energy band, e.g., below 100 keV. Therefore, there is strong evidence for the thermal component in class (i).\vspace{-4pt}
	
\begin{figure}[H]

\begin{adjustwidth}{-\extralength}{0cm}
		\centering
		\subfigure[~BAND]{
			\includegraphics[width=0.65\textwidth]{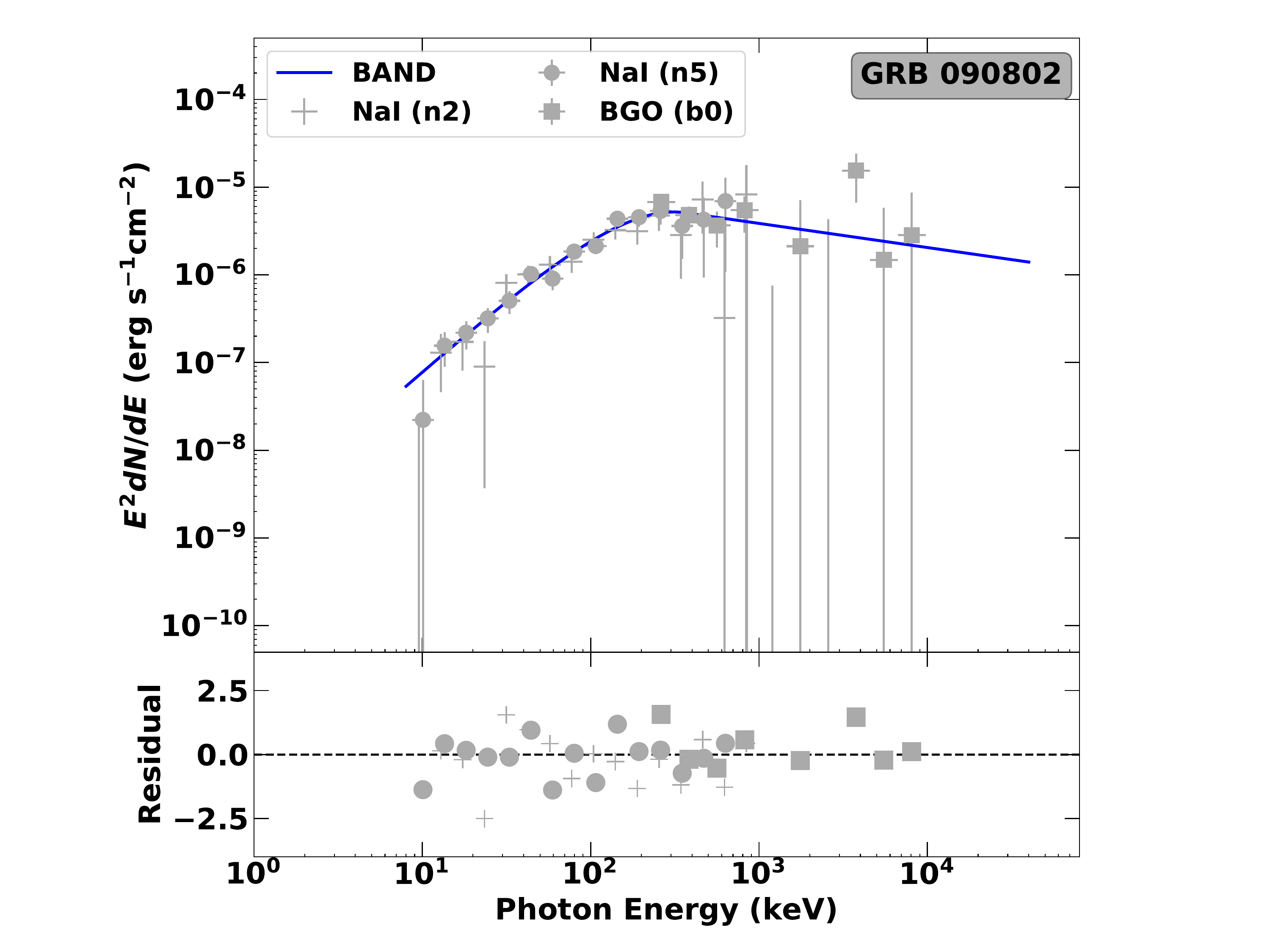}}
		\subfigure[~CPL]{
			\includegraphics[width=0.65\textwidth]{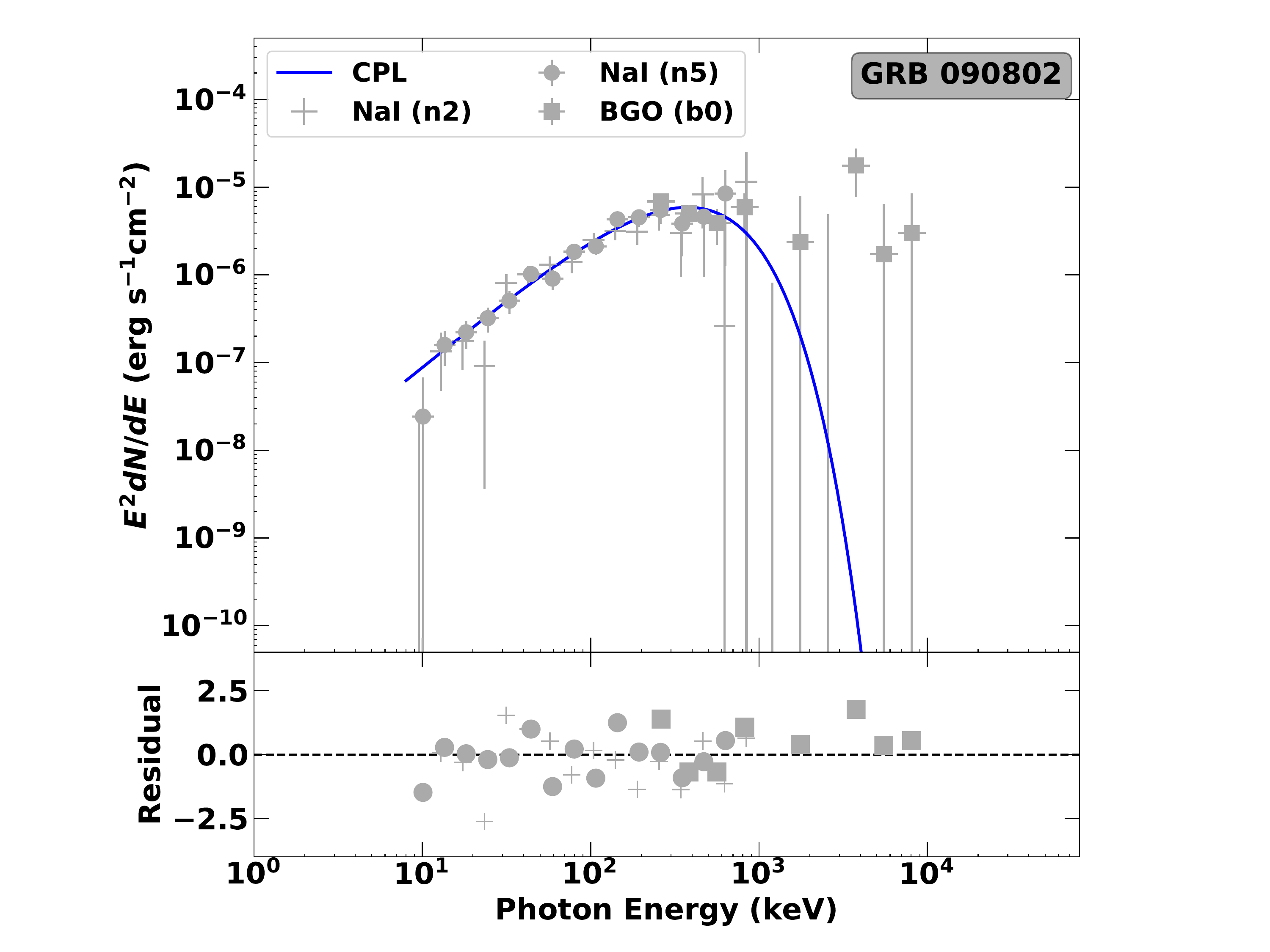}}\\
		\subfigure[~BAND + BB]{
			\includegraphics[width=0.65\textwidth]{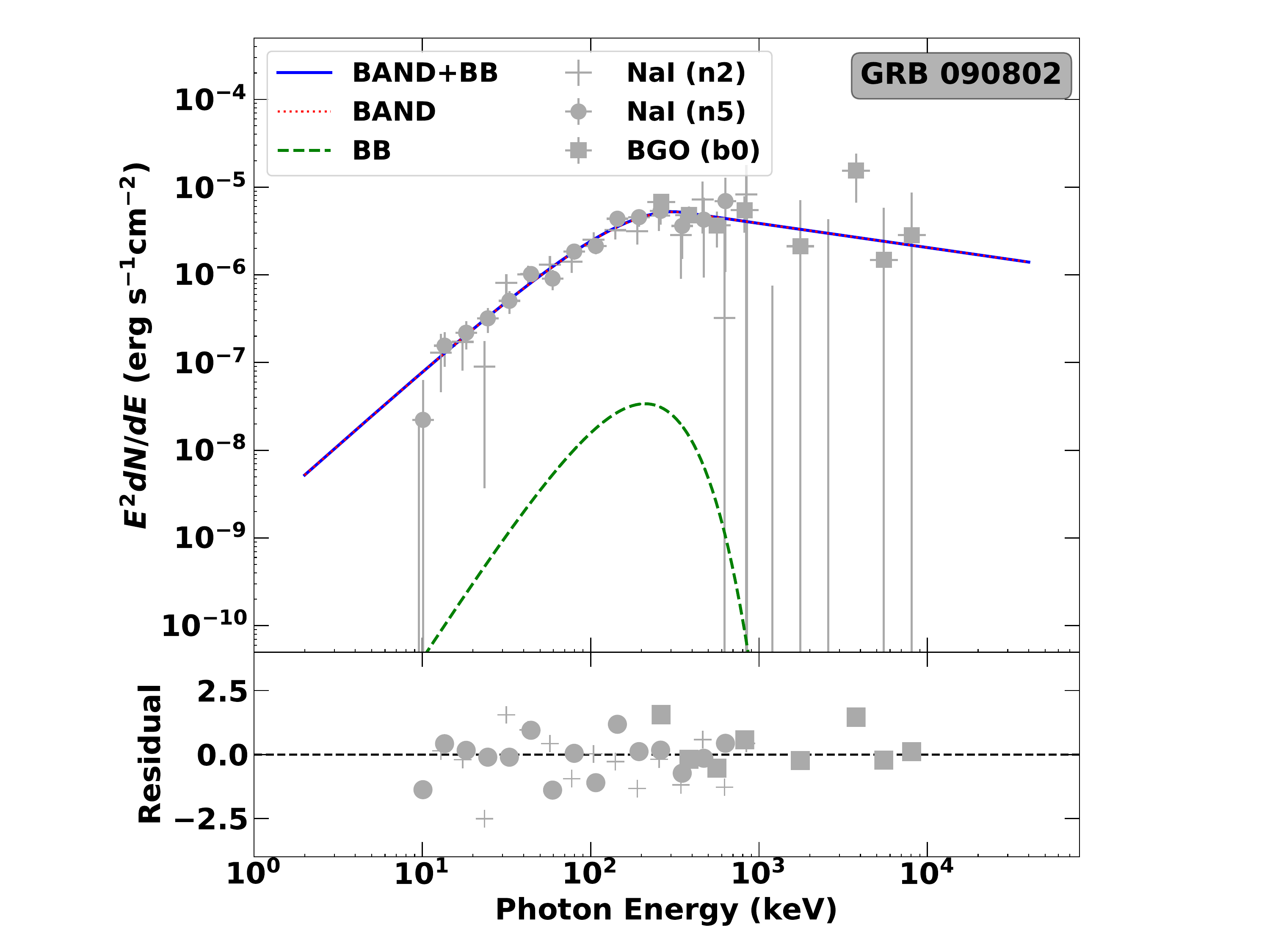}}
		\subfigure[~CPL + BB]{
			\includegraphics[width=0.65\textwidth]{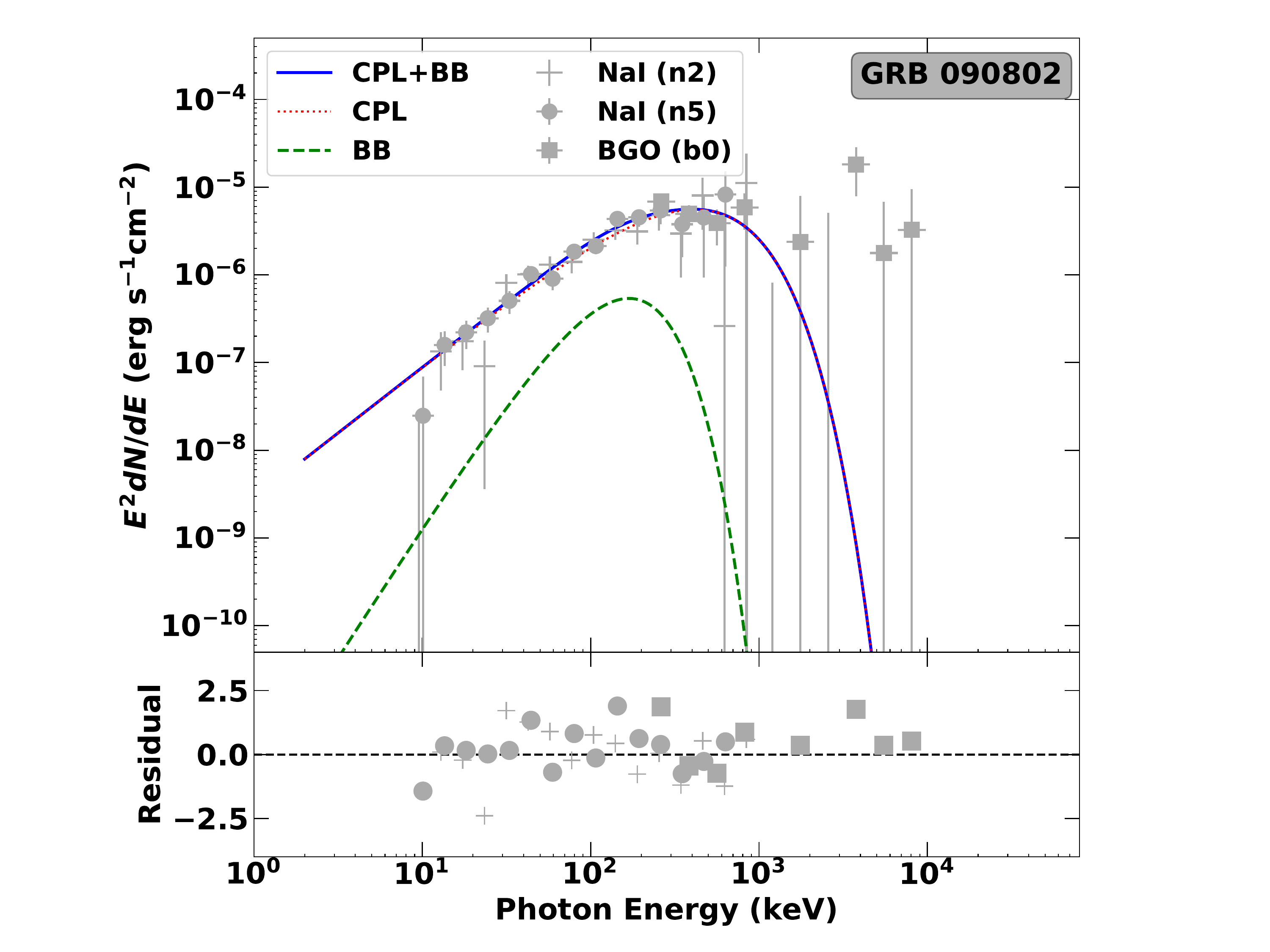}}
\end{adjustwidth}
		\caption{Spectral-fitting results for one GRB in class (ii) that with $R < 0.1$ in both of the \mbox{BAND + BB} model and the CPL + BB model, such as GRB 090802.}
		\label{fig:subsample2}
	\end{figure}

	For class (ii), there are two GRBs included, GRB 090802 and GRB 091126. The results of GRB 090802 are plotted in Figure~\ref{fig:subsample2}. As seen in Figure~\ref{fig:subsample2}, the BB component in both model is exceeded by the non-thermal component, such as the BAND component or CPL component. There is little evidence for a detection of the thermal component in this class of GRBs.
	
	\vspace{-4pt}
	
\begin{figure}[H]

\begin{adjustwidth}{-\extralength}{0cm}
		\centering
		\subfigure[~BAND]{
			\includegraphics[width=0.65\textwidth]{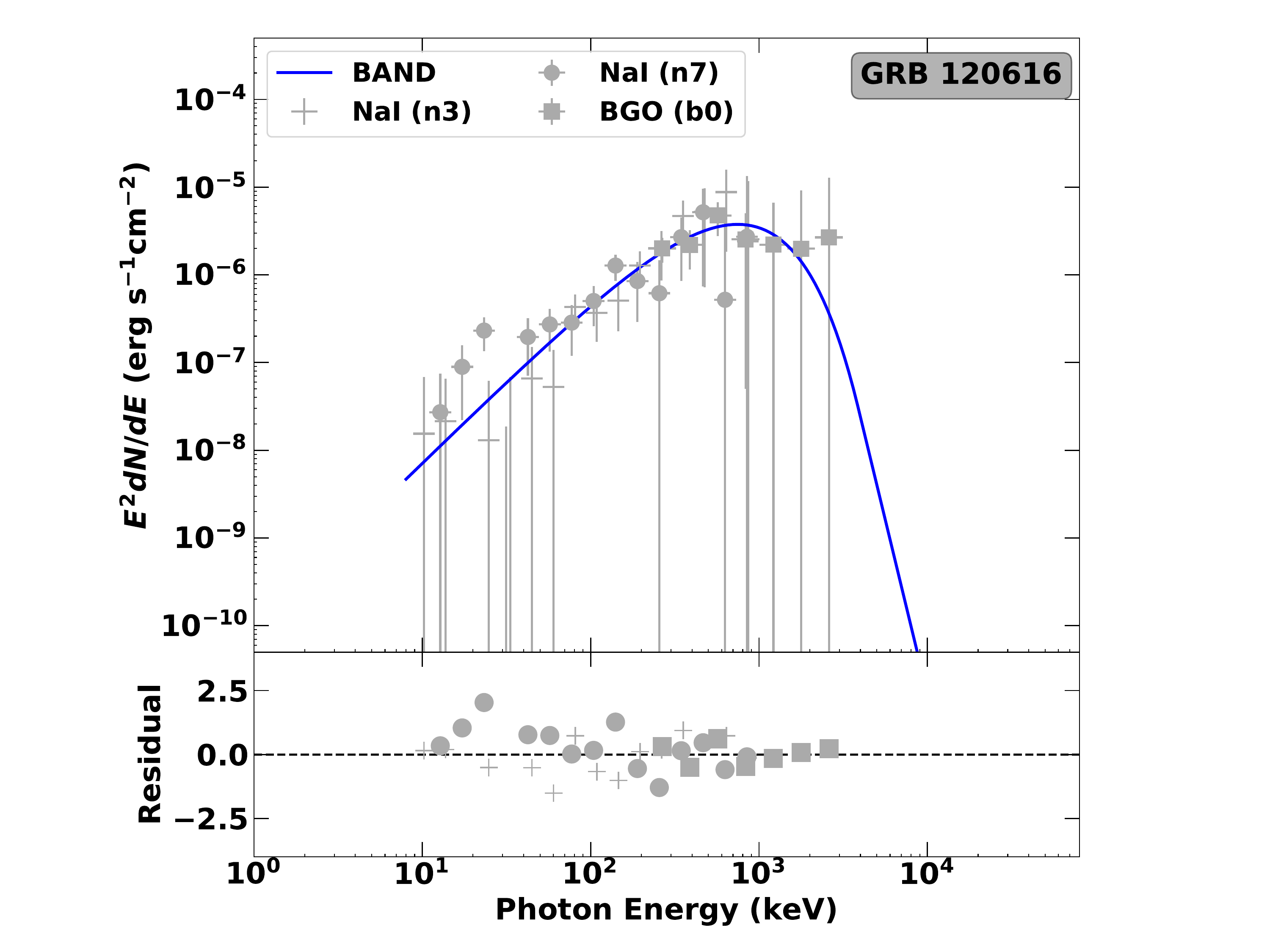}}
		\subfigure[~CPL]{
			\includegraphics[width=0.65\textwidth]{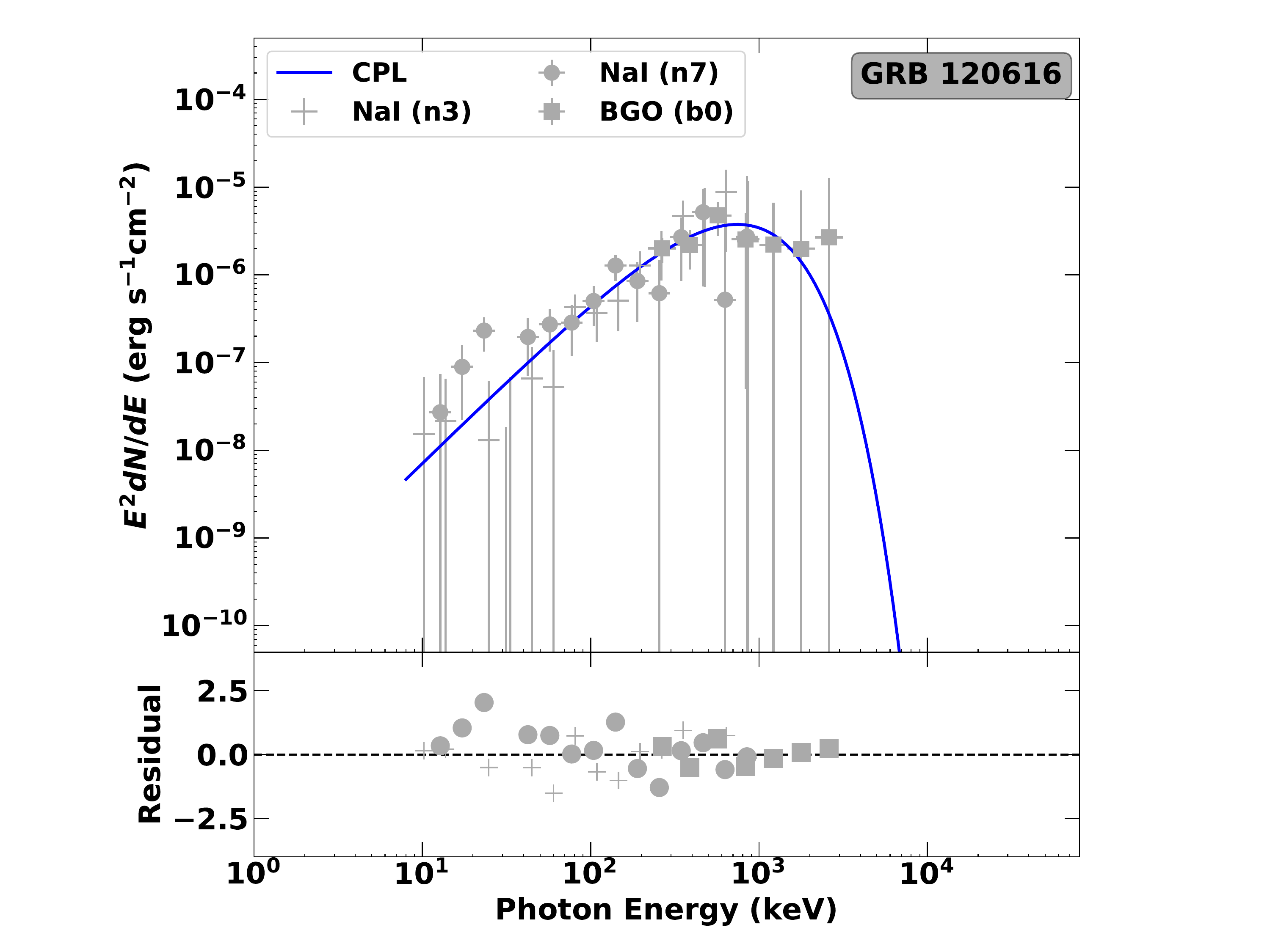}}\\
		\subfigure[~BAND + BB]{
			\includegraphics[width=0.65\textwidth]{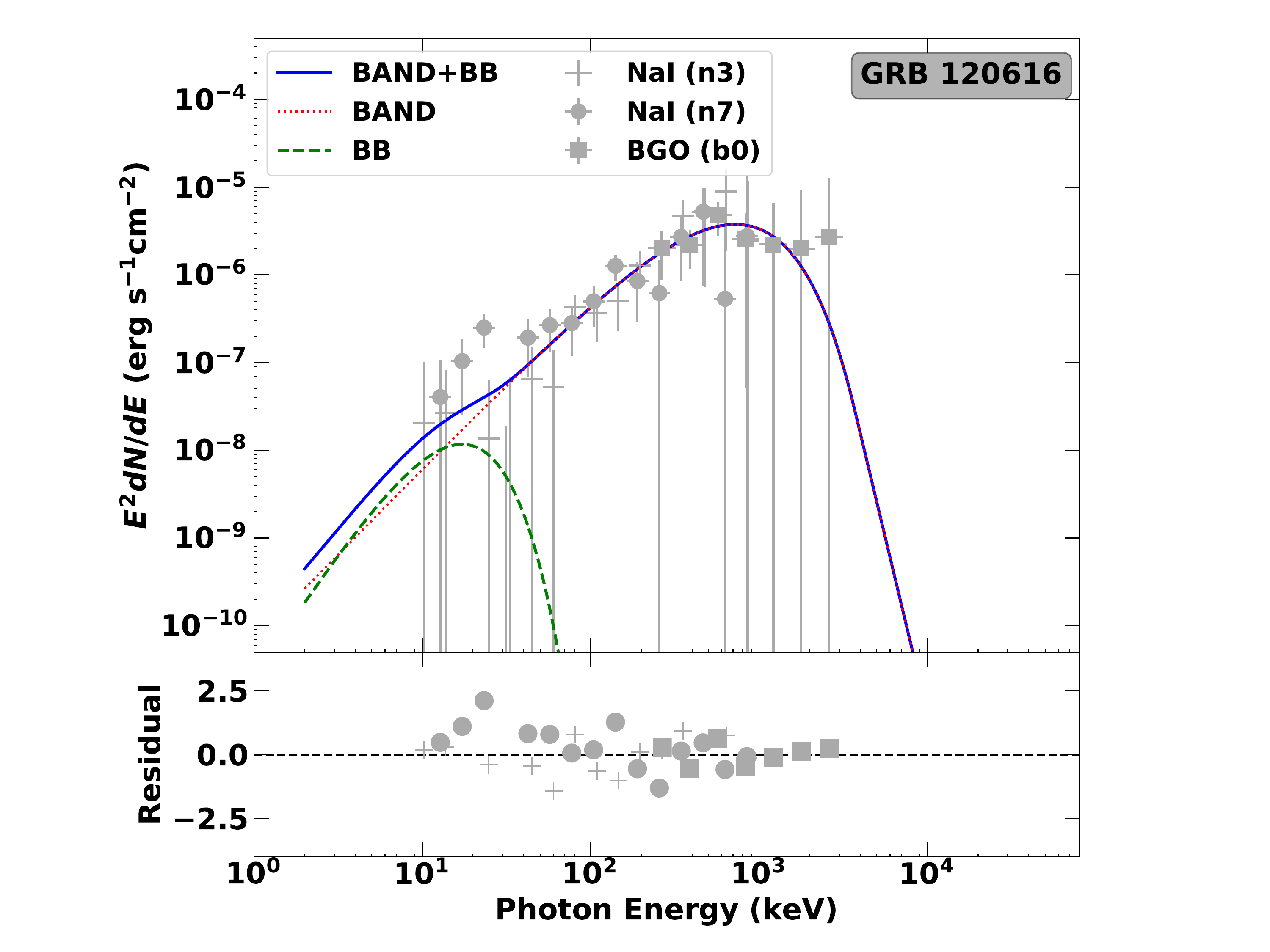}}
		\subfigure[~CPL + BB]{
			\includegraphics[width=0.65\textwidth]{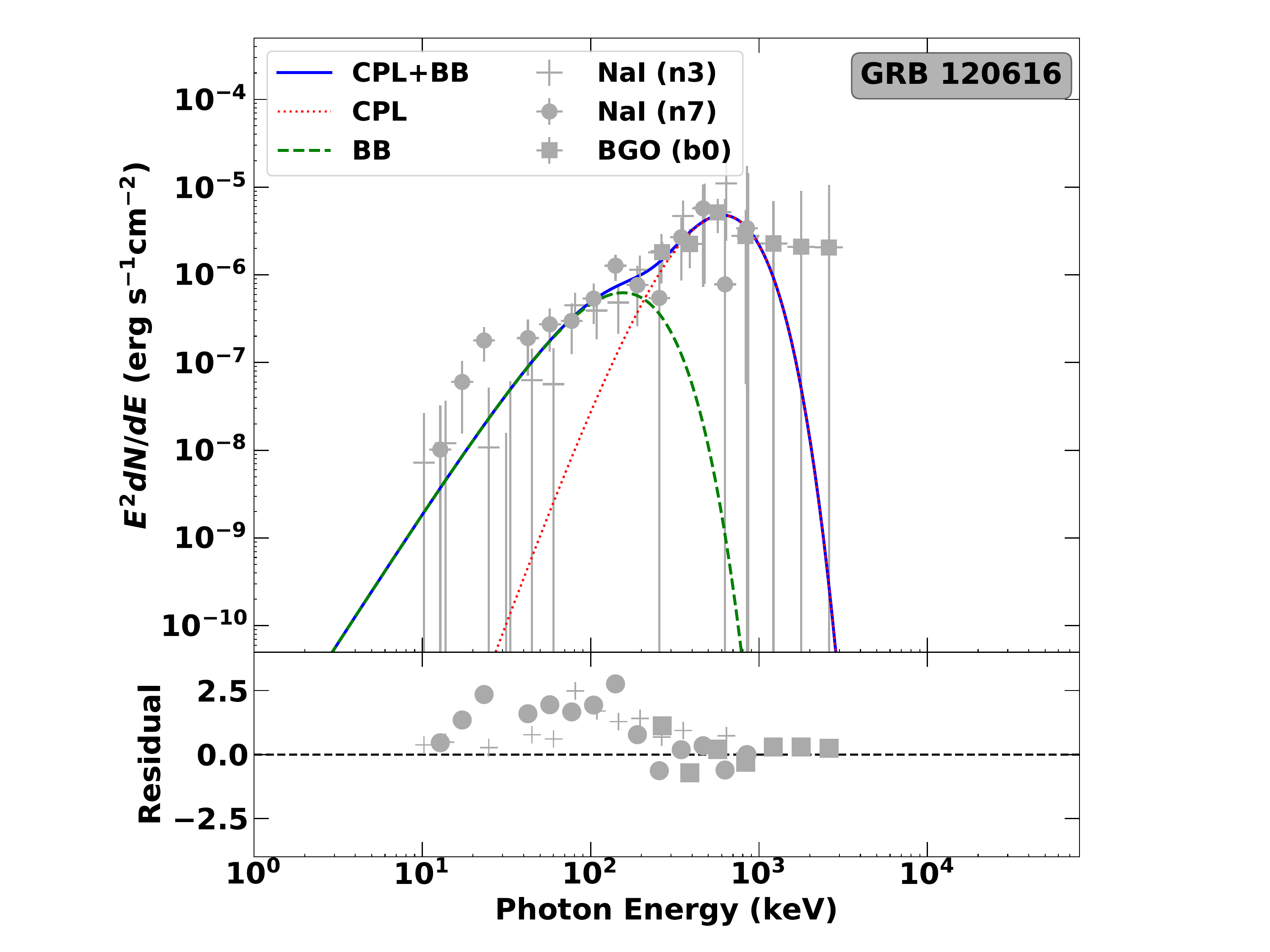}}
\end{adjustwidth}
		\caption{Spectral-fitting results for one GRB in class (iii) that with $R > 0.1$ in one of the \mbox{BAND + BB} model and the CPL + BB model, such as GRB 081221.}
		\label{fig:subsample3}
	\end{figure}

	Five GRBs belong to class (iii), GRB 120616, GRB 160822, GRB 180103, GRB 80602 and GRB 190505. Spectral fitting results of GRB 120616 are plotted in Figure~\ref{fig:subsample3}. As seen in Figure~\ref{fig:subsample3}, the BB component in the BAND + BB model is not strong as the BAND component above about 20 keV. However, in the CPL + BB model, the CPL component has a lower contribution to observations than the BB component. There is moderate evidence for the thermal component detection in class (iii).
	
	As a short conclusion, at least eight GRBs in our sample provide moderate evidence for the detection of the thermal component.
	
	\section{Discussion \label{discussion}}
	
	\subsection{Does a Very Short GRB have a Very Hard Spectrum?}
	Based on the results above, most GRBs in our sample have one hard spectral component, either with a hard low-energy slope or with a hard peak energy. For the hard low-energy slope discovered in the very short GRBs, it might challenge the synchrotron radiation model in the prompt phase, which is also discussed in other works~\cite{Tang2021,Zhao2022a,Zhao2022b}. Alternatively, the physical synchrotron model invoking a decaying magnetic field and/or an evolution of the cooling regime could also reproduce such a hard slope~\cite{Uhm2014,Burgess2020}.
	
	\textls[-10]{A high peak energy ($E_{\rm peak, IS}$) in internal shock model could be achieved if a very short GRB has a very small variability ($\tau$) during the relativistic ejection, such as \mbox{$E_{\rm peak, IS} \propto \tau^{-1}$~\citep{Guiriec2010}.}} In the photospheric model, a high peak energy ($E_{\rm peak, thermal}$) would be obtained if the outflow in a short GRB has systematically ultra-high Lorentz factors~\citep{Guiriec2010}. Our result that a very short GRB usually has a hard spectrum implies that searching for the hard spectral feature in LGRBs is possible, e.g., performing the spectral analysis in the early peakflux period of a LGRB. 
	
	\subsection{Is there a Common Thermal Component in the Very Short Bursts?}
	Although not strongly we claim the detected thermal component, however, the moderate evidence for the thermal-component detection is discovered in eight GRBs. This is also confirmed in several bright GRBs, which have significant detections both in Fermi/GBM and Fermi Large Area Telescope (Fermi/LAT)~\cite{Tang2021}.
	
	However, for most GRBs in our sample, the flux ratio between the BB component and the BAND/CPL component is usually low, which implies that the non-thermal component still dominates the photon flux in the high-energy band, e.g., above 100 keV. In other word, there are rare GRBs with the thermal component dominating the global prompt emission. This is also because we only considered the standard Plank function to fit the thermal component, which corresponds to the case of the none-dissipative photosphere~\citep{Ryde2005,Peer2007,Peer2008}. It is found that such a narrow BB spectral shape could be broadened when considering the case of the subphotospheric dissipation~\citep{Bromberg2011,Levinson2012,Beloborodov2017,Ahlgren2019}. Further theoretical calculations might shed light on the contribution by the thermal component as well as its evolution, which is out of the scope of this work. 
	
	\section{Summary and Conclusions \label{conclusion}}
	Ten very short GRBs detected by Fermi/GBM have been analyzed, whose observed spectra are fitted by four models, the BAND model, the CPL model, the \mbox{BAND + BB} model, and the CPL + BB model. Both the derived low-energy spectral index and the peak energy imply a common hard spectrum feature for all GRBs. We found moderate evidence for the detection of thermal component in eight GRBs, which indicates a possible common feature of these very short GRBs. However, the thermal component in these short GRBs contributes a small proportion of the global prompt emission, especially above 100 keV. The modified thermal-radiation mechanism could enhance the proportion significantly, such as in subphotospheric dissipation. 
	

	
	
	\authorcontributions{Data curation, Ying-Yong Hu; Formal analysis, Zan Zhu; Methodology, Yao-Lin Huang; Project administration, Qing-Wen Tang; Software, Jia-Wei Huang.}
	
	\funding{\textls[-10]{This work is funded by National Nature Science Foundation of China grant numbers 11903017 and 12065017, and Jiangxi Provincial Natural Science Foundation under grant 20212BAB201029.} }
	
	\dataavailability{Data available on reasonable request to the authors.}
	
	\acknowledgments{We appreciate the anonymous referees for the constructive suggestions.}
	
	\conflictsofinterest{The authors declare no conflict of interest.}

	
	
\begin{adjustwidth}{-\extralength}{0cm}
	\printendnotes[custom]{}
	\end{adjustwidth}
	
\begin{adjustwidth}{-\extralength}{0cm}	
	\reftitle{References}
	
	

\begin{thebibliography}{999}
			
			\bibitem[Woosley(1993)]{Woosley1993} Woosley, S.E. \ Gamma-Ray Bursts from Stellar Mass Accretion Disks around Black Holes.\  \emph{Astrophys. J.} \textbf{1993}, \emph{405}, 273.
			
			\bibitem[Paczy{\'n}ski(1998)]{Paczynski1998} Paczy{\'n}ski, B.\ Are Gamma-Ray Bursts in Star-Forming Regions?\ \emph{Astrophys. J.} \textbf{1998}, \emph{494}, L45--L48.
			
			\bibitem[Galama~et~al.(1998)]{Galama1998} Galama, T.J.;  Vreeswijk, P.M.;  van Paradijs, J.;  Kouveliotou, C.;  Augusteijn, T.;  B{\"o}hnhardt, H.;  Brewer, J.P.;  Doublier, V.;  Gonzalez, J.F.;  Leibundgut, B.;~et~al. \ An unusual supernova in the error box of the {\ensuremath{\gamma}}-ray burst of 25 April 1998.\ \emph{Nature} \textbf{1998}, \emph{395}, 670--672.
			
			\bibitem[Stanek~et~al.(2003)]{Stanek2003} Stanek, K.Z.;   Matheson, T.;  Garnavich, P.M.;  Martini, P.;  Berlind, P.;  Caldwell, N.;  Challis, P.;  Brown, W.R.;  Schild, R.;  Krisciunas, K.;~et~al. \ Spectroscopic Discovery of the Supernova 2003dh Associated with GRB 030329.\ \emph{Astrophys. J.} \textbf{2003}, \emph{591}, L17--L20.
			
			\bibitem[Woosley and Bloom(2006)]{Woosley2006} Woosley, S.E.; Bloom, J.S. \ The Supernova Gamma-Ray Burst Connection.\ \emph{Annu. Rev. Astron. Astrophys.} \textbf{2006}, \emph{44}, 507--556. 
			
			\bibitem[Fruchter~et~al.(2006)]{Fruchter2006} Fruchter, A.S.;   Levan, A.J.;  Strolger, L.;  Vreeswijk, P.M.;  Thorsett, S.E.;  Bersier, D.;  Burud, I.;  Castro Cerón, J.M.;  Castro-Tirado, A.J.;  Conselice, C.;~et~al.\ Long {\ensuremath{\gamma}}-ray bursts and core-collapse supernovae have different environments.\ \emph{Nature} \textbf{2006}, \emph{441}, 463--468.
			
			\bibitem[Abbott~et~al.(2017)]{Abbott2017} Abbott, B.P.;  Abbott, R.;  Abbott, T.D.;  Acernese, F.;  Ackley, K.;  Adams, C.;  Adams, T.;  Addesso, P.;  Adhikari, R.X.;  Adya, V.B.;~et~al.\ Gravitational Waves and Gamma-Rays from a Binary Neutron Star Merger: GW170817 and GRB 170817A.\ \emph{Astrophys. J. Lett.} \textbf{2017}, \emph{848}, L13.
			
			\bibitem[Abdo~et~al.(2009)]{Abdo2009} Abdo, A.A.;  Ackermann, A.;  Ajello, M.;  Asano, K.;  Atwood, W.B.;  Axelsson, M.;  Baldini, L.;  Ballet, J.;  Barbiellini, G.;  Baring, M.G.;~et~al.\ Fermi Observations of GRB 090902B: A Distinct Spectral Component in the Prompt and Delayed Emission.\ \emph{Astrophys. J. Lett.} \textbf{2009}, \emph{706}, L138--L144.
			
			\bibitem[Guiriec~et~al.(2013)]{Guiriec2013} Guiriec, S.;  Daigne, F.;  Hascoët, R.;  Vianello, G.;  Ryde, F.;  Mochkovitch, R.;  Kouveliotou, C.;  Xiong, S.;  Bhat, P.N.;  Foley, S.;~et~al.\ Evidence for a Photospheric Component in the Prompt Emission of the Short GRB 120323A and Its Effects on the GRB Hardness-Luminosity Relation.\ \emph{Astrophys. J.} \textbf{2013}, \emph{770}, 32.
			
			\bibitem[Tang~et~al.(2021)]{Tang2021} Tang, Q.-W.;  Wang, K.;  Li, L.;  Liu, R.-Y. \ Prevalence of Extra Power-Law Spectral Components in Short Gamma-Ray Bursts.\ \emph{Astrophys. J.} \textbf{2021}, \emph{922}, 255.
			
			\bibitem[Zhao \& Tang(2022)]{Zhao2022a} Zhao, P.-W.;  Tang, Q.-W. \ A Comprehensive Study of Bright Fermi-GBM Short Gamma-ray Bursts: I. Multi-Pulse Lightcurves and Multi-Component Spectra.\ \emph{Universe} \textbf{2022}, \emph{8}, 159.
			
			\bibitem[Zhao~et~al.(2022)]{Zhao2022b} Zhao, P.-W.;  Tang, Q.-W.;  Zou, Y.-C.;  Wang, K.\ Detection of a Prompt Fast-variable Thermal Component in the Multipulse Short Gamma-Ray Burst 170206A.\ \emph{Astrophys. J.} \textbf{2022}, \emph{929}, 179.
			
			\bibitem[Arimoto~et~al.(2016)]{Arimoto2016} Arimoto, M.;  Asano, K.;  Ohno, M.;  Veres, P.;  Axelsson, M.;  Bissaldi, E.;  Tachibana, Y.;  Kawai, N. \ High-energy Non-thermal and Thermal Emission from GRB141207A Detected by Fermi.\ \emph{Astrophys. J.} \textbf{2016}, \emph{833}, 139.
			
			\bibitem[L{\"u}~et~al.(2017)]{Lv2017} L{\"u}, H.-J.;  L{\"u}, J.;  Zhong, S.-Q.;   Huang, X.-L.;  Zhang, H.-M.;  Lan, L.;  Xie, W.;  Lu, R.-J.;  Liang, E.-W.\ Extremely Bright GRB 160625B with Multiple Emission Episodes: Evidence for Long-term Ejecta Evolution.\ \emph{Astrophys. J.} \textbf{2017}, \emph{849}, 71.
			
			\bibitem[Hou~et~al.(2018)]{Hou2018} Hou, S.-J.;  Zhang, B.-B.;  Meng, Y.-Z.;  Wu, X.-F.;  Liang, E.-W.;  L{\"u}, H.-J.;  Liu, T.;  Liang, Y.-F.;  Lin, L.;  Lu, R.-J.;~et~al.\ Multicolor Blackbody Emission in GRB 081221.\ \emph{Astrophys. J.} \textbf{2018}, \emph{866}, 13.
			
			\bibitem[Iyyani and Sharma(2021)]{Iyyani2021} Iyyani, S.;  Sharma, V. \ Study of the Prompt Emission of Short Gamma-Ray Bursts Using a Multicolor Blackbody: A Clue to the Viewing Angle.\ \emph{Astrophys. J. Suppl. Ser.} \textbf{2021}, \emph{255}, 25.
			
			\bibitem[Ahlgren~et~al.(2015)]{Ahlgren2015} Ahlgren, B.;  Larsson, J.;  Nymark, T.;  Ryde, F.;  Pe'er, A. \ Confronting GRB prompt emission with a model for subphotospheric dissipation.\ \emph{Mon. Not. R. Astron. Soc. Lett.} \textbf{2015}, \emph{454}, L31--L35.
			
			\bibitem[Beloborodov(2017)]{Beloborodov2017} Beloborodov, A.M. \ Sub-photospheric Shocks in Relativistic Explosions.\ \emph{Astrophys. J.} \textbf{2017}, \emph{838}, 125.
			
			\bibitem[Ryde~et~al.(2019)]{Ryde2019} Ryde, F.;  Yu, H.-F.;  Dereli-B{\'e}gu{\'e}, H.;  Lundman, C.;  Pe'er, A.;  Li, L. \ On the $\alpha$-intensity correlation in gamma-ray bursts: Subphotospheric heating with varying entropy.\ \emph{Mon. Not. R. Astron. Soc. Lett.} \textbf{2019}, \emph{485}, 474--497.
			
			\bibitem[Lundman~et~al.(2013)]{Lundman2013} Lundman, C.;  Pe'er, A.;  Ryde, F. \ A theory of photospheric emission from relativistic, collimated outflows.\ \emph{Mon. Not. R. Astron. Soc. Lett.} \textbf{2013}, \emph{428}, 2430--2442.
			
			\bibitem[Meng~et~al.(2019)]{Meng2019} Meng, Y.-Z.;  Liu, L.-D.;  Wei, J.-J.;  Wu, X.-F.;  Zhang, B.-B. \ The Time-resolved Spectra of Photospheric Emission from a Structured Jet for Gamma-Ray Bursts.\ \emph{Astrophys. J.} \textbf{2019}, \emph{882}, 29.
			
			\bibitem[Meng~et~al.(2022)]{Meng2022} Meng, Y.-Z.;  Geng, J.-J.;  Wu, X.-F. \ The photosphere emission spectrum of hybrid relativistic outflow for gamma-ray bursts.\ \emph{Mon. Not. R. Astron. Soc. Lett.} \textbf{2022}, \emph{509}, 6047--6058.
			
			\bibitem[Lazzati \& Begelman(2010)]{Lazzati2010} Lazzati, D.;  Begelman, M.C. \ Non-thermal Emission from the Photospheres of Gamma-ray Burst Outflows. I. High-Frequency Tails.\ \emph{Astrophys. J.} \textbf{2010}, \emph{725}, 1137--1145.
			
			\bibitem[Giannios(2012)]{Giannios2012} Giannios, G. \ The peak energy of dissipative gamma-ray burst photospheres.\ \emph{Mon. Not. R. Astron. Soc. Lett.} \textbf{2012}, \emph{422}, 3092--3098.
			
			\bibitem[Vyas~et~al.(2021)]{Vyas2021a} Vyas, M.;  Pe'er, A.;  Eichler, D. \ A Backscattering-dominated Prompt Emission Model for the Prompt Phase of Gamma-Ray Bursts.\ \emph{Astrophys. J.} \textbf{2021}, \emph{908}, 9.
			
			\bibitem[Vyas~et~al.(2021)]{Vyas2021b} Vyas, M.;  Pe'er, A.;  Eichler, D. \ Predicting Spectral Parameters in the Backscattering-dominated Model for the Prompt Phase of GRBs.\ \emph{Astrophys. J. Lett.} \textbf{2021}, \emph{918}, L12.
			
			\bibitem[Guiriec~et~al.(2015)]{Guiriec2015} Guiriec, S.;  Kouveliotou, C.;  Daigne, F.;  Zhang, B.;  Hasco{\"e}t, R.;  Nemmen, R.-S.;  Thompson, D.-J.;  Bhat, P.-N.;  Gehrels, N.;  Gonzalez, M.-M.;~et~al.\ Toward a Better Understanding of the GRB Phenomenon: A New Model for GRB Prompt Emission and its Effects on the New L$_{i}$$^{NT}$- E$_{peak,i}$$^{rest,NT}$ Relation.\ \emph{Astrophys. J.} \textbf{2015}, \emph{807}, 148. 
			
			\bibitem[Guiriec~et~al.(2010)]{Guiriec2010} Guiriec, S.;  Briggs, M.S.;  Connaugthon, V.;  Kara, E.;  Daigne, F.;  Kouveliotou, C.;  van der Horst, A.J.;  Paciesas, W.;  Charles A.M.;  Bhat, P.N.;~et~al.\ Time-resolved Spectroscopy of the Three Brightest and Hardest Short Gamma-ray Bursts Observed with the Fermi Gamma-ray Burst Monitor.\ \emph{Astrophys. J.} \textbf{2010}, \emph{725}, 225.
			
			\bibitem[Axelsson~et~al.(2012)]{Axelsson2012} Axelsson, M.;  Baldini, L.;  Barbiellini, G.;  Baring, M.G.;  Bellazzini, R.;  Bregeon, J.;  Brigida, M.;  Bruel, P.;  Buehler, R.;~Caliandro, G.A. \ GRB110721A: An Extreme Peak Energy and Signatures of the Photosphere.\ \emph{Astrophys. J. Lett.} \textbf{2012}, \emph{757}, 31.
			
			\bibitem[Burgess~et~al.(2014)]{Burgess2014} Burgess, J.M.;   Preece, R.D.;  Ryde, F.;  Veres, P.;  M{\'e}sz{\'a}ros, P.;  Connaughton, V.;  Briggs, M.;  Pe'er, A.;  Iyyani, S.;  Goldstein, A.;~et~al.\ An Observed Correlation between Thermal and Non-thermal Emission in Gamma-Ray Bursts.\ \emph{Astrophys. J.} \textbf{2014}, \emph{784}, 43.
			
			\bibitem[Yu~et~al.(2019)]{Yu2019} Yu, H.-F.;  Dereli-B{\'e}gu{\'e}, H.;  Ryde, F. \ Bayesian Time-resolved Spectroscopy of GRB Pulses.\ \emph{Astrophys. J.} \textbf{2019}, \emph{886}, 20.
			
			\bibitem[Meegan~et~al.(2009)]{Meegan2009} Meegan, C.;  Lichti, G.;  Bhat, P. N; Bissaldi, E.;  Briggs, M.S.;  Connaughton, V.; Diehl, R.;  Fishman, G.;  Greiner, J.;  Hoover, A.S.;~et~al.\ The Fermi Gamma-ray Burst Monitor.\ \emph{Astrophys. J.} \textbf{2009}, \emph{702}, 791--804. 
			
			\bibitem[Ahlgren~et~al.(2019)]{Ahlgren2019} Ahlgren, B.;  Larsson, J.;  Ahlberg, E.;  Lundman, C.;  Ryde, F.;  Pe'er, A. \ Testing a model for subphotospheric dissipation in GRBs: Fits to Fermi data constrain the dissipation scenario.\ \emph{Mon. Not. R. Astron. Soc. Lett.} \textbf{2019}, \emph{485}, 474--497.


			\bibitem[Band~et~al.(1993)]{Band1993} Band, D.;  Matteson, J.;  Ford, L.;  Schaefer, B.;  Palmer, D.;  Teegarden, B.;  Cline, T.;  Briggs, M.;  Paciesas, W.;  Pendleton, G.;~et~al.\ BATSE Observations of Gamma-Ray Burst Spectra. I. Spectral Diversity.\ \emph{Astrophys. J.} \textbf{1993}, \emph{413}, 281.

			
			\bibitem[Norris~et~al.(2005)]{Norris2005} Norris, J.P.;  Bonnell, J.T.;  Kazanas, D.;  Scargle, J.D.;  Hakkila, J.;  Giblin, T.W.\ Long-Lag, Wide-Pulse Gamma-Ray Bursts.\ \emph{Astrophys. J.} \textbf{2005}, \emph{627}, 324.


			\bibitem[Vianello~et~al.(2015)]{Vianello2015} Vianello, G.;  Lauer, Robert J.;  Younk, P.;  Tibaldo, L.;  Burgess, J.M.;  Ayala, H.;  Harding, P.;  Hui, M.;  Omodei, N.;  Zhou, H.\ The Multi-Mission Maximum Likelihood framework (3ML).\  \emph{arXiv} \textbf{2015}, arXiv:1507.08343.

			
			\bibitem[Akaike(1974)]{Akaike1974} Akaike, H.\ A New Look at the Statistical Model Identification. \emph{IEEE Trans. Autom. Control.} \textbf{1974}, \emph{19}, 716.
			
			\bibitem[Schwarz(1978)]{Schwarz1978} Schwarz, G.\ Estimating the Dimension of a Model.\ \emph{Ann. Stat.} \textbf{1978}, \emph{6}, 461.

			\bibitem[von Kienlin~et~al.(2020)]{vonKienlin2020} von Kienlin, A.;  Meegan, C.A.;  Paciesas, W.S.;  Bhat, P.N.;  Bissaldi, E.;  Briggs, M.S.;  Burns, E.;  Cleveland, W.H.;  Gibby, M.H.;  Giles, M.M.;~et~al.\ The Fourth Fermi-GBM Gamma-Ray Burst Catalog: A Decade of Data.\ \emph{Astrophys. J.} \textbf{2020}, \emph{893}, 46.



			

			



			

			
			\bibitem[Kouveliotou~et~al.(1993)]{Kouveliotou1993} Kouveliotou, C.;  Meegan, C.A.;  Fishman, G.J.;  Bhat, N.P.;  Briggs, M.S.;  Koshut, T.M.;  Paciesas, W.S.;  Pendleton, G.N.\ Identification of Two Classes of Gamma-Ray Bursts.\ \emph{Astrophys. J. Lett.} \textbf{1993}, \emph{413}, L101. 
			
			\bibitem[Zhang~et~al.(2009)]{Zhang2009} Zhang, B.;  Zhang, B.B.;  Virgili, F.J.;  Liang, E.W.;  Kann, D.A.;  Wu, X.F.;  Proga, D.;  Lv, H.J.;  Toma, K.;  M{\'e}sz{\'a}ros, P.;~et~al.\ Discerning the Physical Origins of Cosmological Gamma-ray Bursts Based on Multiple Observational Criteria: The Cases of z = 6.7 GRB 080913, z = 8.2 GRB 090423, and Some Short/Hard GRBs.\ \emph{Astrophys. J.} \textbf{2009}, \emph{703}, 1696--1724.
			
			\bibitem[Burns~et~al.(2018)]{Burns2018} Burns, E.;  Veres, P.;  Connaughton, V.;  Racusin, J.;  Briggs, M.S.;  Christensen, N.;  Goldstein, A.;  Hamburg, R.;  Kocevski, D.;  McEnery, J.;~et~al.\ Fermi GBM Observations of GRB 150101B: A Second Nearby Event with a Short Hard Spike and a Soft Tail.\ \emph{Astrophys. J.} \textbf{2018}, \emph{863}, 34.
			
			\bibitem[Sari~et~al.(1998)]{Sari1998} Sari, R.;  Piran, T.;  Narayan, R.\ Spectra and Light Curves of Gamma-Ray Burst Afterglows.\ \emph{Astrophys. J.} \textbf{1998}, \emph{497}, L17--L20.
			
			\bibitem[Uhm \& Zhang(2014)]{Uhm2014} Uhm, Z.L.; Zhang, B.\ Fast-cooling synchrotron radiation in a decaying magnetic field and $\gamma$-ray burst emission mechanism.\ \emph{Nat. Phys.} \textbf{2014}, \emph{10}, 351.
			
			\bibitem[Burgess~et~al.(2020)]{Burgess2020} Burgess, J.M.;  B{\'e}gu{\'e}, D.;  Greiner, J.;  Giannios,
			D.;  Bacelj, A.;  Berlato, F.\ \emph{Nat. Astron.} \textbf{2020}, \emph{4}, 174.
			
			\bibitem[Ryde(2005)]{Ryde2005} Ryde, F. \ Is Thermal Emission in Gamma-Ray Bursts Ubiquitous?.\ \emph{Astrophys. J.} \textbf{2005}, \emph{625}, L95--L98.
			
			\bibitem[Pe'er~et~al.(2007)]{Peer2007} Pe'er, A.;  Ryde, F.;  Wijers, R.A.M.J.;  M{\'e}sz{\'a}ros, P.;  Rees, M.J. \ A New Method of Determining the Initial Size and Lorentz Factor of Gamma-Ray Burst Fireballs Using a Thermal Emission Component.\ \emph{Astrophys. J.} \textbf{2007}, \emph{664}, L1--L4.
			
			\bibitem[Pe'er(2008)]{Peer2008} Pe'er, A. \ Temporal Evolution of Thermal Emission from Relativistically Expanding Plasma.\ \emph{Astrophys. J.} \textbf{2008}, \emph{682}, 463--473.
			
			\bibitem[Bromberg~et~al.(2011)]{Bromberg2011} Bromberg, O.;  Mikolitzky, Z.;  Levinson, A. \ Sub-photospheric Emission from Relativistic Radiation Mediated Shocks in GRBs.\ \emph{Astrophys. J.} \textbf{2011}, \emph{733}, 85.
			
			\bibitem[Levinson(2012)]{Levinson2012} Levinson, A. \ Observational Signatures of Sub-photospheric Radiation-mediated Shocks in the Prompt Phase of Gamma-Ray Bursts.\ \emph{Astrophys. J.} \textbf{2012}, \emph{756}, 174.
			
			
			
			
		\end{thebibliography}
	

		
	\end{adjustwidth}
	
\end{document}